\documentclass[12pt,preprint]{aastex}

\shorttitle{Diffuse Continua}
\shortauthors{Korista \& Goad}
\begin{document}

\title{The Variable Diffuse Continuum Emission of Broad Line Clouds}

\author{Kirk T.\ Korista\altaffilmark{1}}
\affil{Department of Physics, Western Michigan University}
\affil{Kalamazoo, MI 49008-5252}
\email{korista@wmich.edu}

\and

\author{Michael R.\ Goad\altaffilmark{2}}
\affil{Department of Physics and Astronomy, University of Leicester}
\affil{University Road, Leicester, LE1 7RH, England, UK}
\email{mrg@star.le.ac.uk}

\begin{abstract}
We investigate the wavelength-dependent intensity and reverberation
properties of the UV-optical diffuse continuum emission expected from
broad emission line gas. The ``locally optimally-emitting clouds''
picture is adopted, with the cloud distribution functions in gas
density and distance from the ionizing source determined by a fit to
the mean UV emission line spectrum from the 1993 {\em HST} campaign of
NGC~5548 in our previous paper. The model Balmer continuum's strength
and variability characteristics are in agreement with those derived
from the observations of NGC~5548. A key prediction is a
wavelength-dependent lag across the UV-optical spectrum which can
broadly mimic the signature from X-ray reprocessing in an accretion
disk, calling into question the discovery claims of the latter in
NGC~7469. The influence of the diffuse continuum on the optical
continuum may result in a small yet significant underestimation of the
characteristic sizes of the regions emitting the optical emission
lines. Its contribution can also alter the inferred spectral energy
distribution of the UV-optical continuum, even well outside the
spectral region near the well-known Balmer jump. The reverberation of
the diffuse continuum emitted by the BLR may account for perhaps
one-third of the observed effect that the $\lambda$1350 --
$\lambda$5100 continuum becomes bluer as it becomes brighter (even
after accounting for non-variable optical starlight).  And as in other
models of the broad emission line gas, a significant yet unobserved
Lyman jump is predicted. We highlight the importance of careful studies
of the UV-optical continuum, especially the Balmer continuum.
\end{abstract}


\objectname[]{NGC~5548}

\normalsize

\section{INTRODUCTION}

From the early work of Davidson \& Netzer (1979) to the more advanced
models of Rees, Netzer, \& Ferland (1989), Goad, Gondhalekar, \&
O'Brien (1993), Kaspi \& Netzer (1999) and many others, much attention
has been given to the theoretical modeling of the prominent broad
emission lines of active galactic nuclei (AGN; including quasars) and
their variability in order to understand the lines' origin and the
story of the AGN phenomenon.  However, with few exceptions (e.g.,
Davidson 1976; Wamsteker et al.\ 1990; Maoz et al.\ 1993) very little
has been said about the diffuse continuum emission from the same
emitters of the broad lines.  Korista et al.\ (1995; hereafter K95)
presented six high signal-to-noise ``continuum window'' light curves of
NGC~5548 between 1145~\AA\/ and 2195~\AA\/ (rest frame; their
Figure~6), and they noted systematic decreases in the variability
amplitude and sharpness of the variations. While they could not measure
it directly, K95 suspected these characteristics to be the signature of
a wavelength-dependent continuum lag, whose origin was either with the
continuum source and/or with contaminating diffuse continuum emission
from the broad line region (BLR). If the diffuse continuum emitted by
broad line clouds is significant, it will have the following impacts on
models for the broad emission line region and the UV-optical continuum
emitter: (1) the lag between the emission line variations and those of
the local ``continuum'' may underestimate the characteristic sizes of
the line emitting regions, (2) the shape of the underlying nuclear
continuum may be misinferred and thus misinterpreted, (3) interband
continuum lags may be misinterpreted, (4) the reverberation of the
diffuse continuum emitted by the BLR may account for perhaps one-third
of the observed effect that the $\lambda$1350 -- $\lambda$5100
continuum becomes bluer as it becomes brighter (even after accounting
for non-variable optical starlight), and (5) as in other models of the
broad emission line gas, a significant yet unobserved Lyman jump is
predicted. Here we investigate the importance of the diffuse continuum
to the mean and variable light spectra of AGN using the well-studied
AGN NGC~5548 as a case in point.

\section{DIFFUSE CONTINUUM EMISSION FROM THE BROAD LINE REGION}

\subsection{The Photoionization Grid}

We adopted the grid of photoionization computations presented in
Korista \& Goad (2000; hereafter, Paper~I) to model the UV broad line
emission from NGC~5548. The reader is directed there for details; here
we mention some of the salient features of the grid.

Using Ferland's spectral synthesis code, {\sc Cloudy} (v90.04) (Ferland
1997; Ferland et al.\ 1998) we generated a grid of photoionization
models of BEL emitting entities, here assumed to be simple slabs, each
of which we assumed has constant gas density and a clear view to the
source of ionizing photons. The grid dimensions spanned 7 orders of
magnitude in total hydrogen gas number density, $7 \le \log~n_H
(\rm{cm}^{-3}) \le 14$, and hydrogen-ionizing photon flux, $17 \le
\log~\Phi_H (\rm{cm^{-2}~s^{-1}}) \le 24$, and stepped in 0.125 decade
intervals in each dimension (3,249 separate {\sc Cloudy} models). We
will refer to the plane defined by these two parameters as the density
-- flux plane.  For the present simulations we assumed all clouds have
a single total hydrogen column density, $N_H = 10^{23}~\rm{cm^{-2}}$.
For very low values of the ionization parameter, $U_H \equiv
\Phi\/_H/n_Hc$, the cloud computations stopped at an electron
temperature of 4000~K rather than at the total hydrogen column density
$10^{23}$~cm$^{-2}$. This artifact had little impact upon the results
presented here (or in Paper~I), since these clouds contributed little
to the light of the model BLR.  Descriptions of the slightly sub-solar
abundances, the incident continuum spectral energy distribution (SED),
and the cloud distribution functions in gas density and radius to model
the mean UV spectrum of NGC~5548 are given in Paper~I.

For the choice of cosmological parameters from Paper~I ($H_{\rm{o}} =
75~\rm{km~s^{-1}~Mpc^{-1}}$; $q_{\rm{o}} = 0.5$; $z = 0.0172$), $\log
\lambda\/L_{\lambda\/1350} \approx 43.54$ ergs~s$^{-1}$ from the mean,
dereddened value of $\lambda\/F_{\lambda}(1350)$ of the 1993 {\em HST}
campaign. The central continuum SED adopted in Paper~I (see also Walter
et al.\ 1994) then finds the mean hydrogen-ionizing luminosity to be
$\log L_{ion} \approx 44.26$ ergs~s$^{-1}$. At this luminosity a $\log
\Phi_H = 20$ $\rm{cm^{-2}~s^{-1}}$ corresponds to a distance from the
continuum source of $R \approx\/~12.6~(75/H_{\rm{o}})$ light-days.

\subsection{The Diffuse Continuous Emission from Broad-Line Clouds}

The diffuse continuum emission includes a component of reflected
incident continuum (see Korista \& Ferland 1998) in addition to true
(i.e., thermal) diffuse, non-line emission from the cloud. The ratios
of twelve diffuse continuum band fluxes to the incident continuum flux
measured at 1215~\AA\/,
$\lambda\/F_{\lambda}(d)/\lambda\/F_{\lambda\/1215}(i)$, spanning the
UV-optical spectrum, are plotted in contour in the density -- flux
plane in Figures~1a,b. Lines of constant ionization parameter run at
$45\degr\/$ angles from the lower left to the upper right of each
panel. In all cases the solid triangle marks the peak ratio. As noted
by Rees, Netzer, \& Ferland (1989), the high density clouds are
efficient emitters of continua.

Across most of the optical-UV spectral region, the diffuse continuum is
dominated by hydrogen free-bound emission. The diffuse to incident
continuum ratio contours generally run gently downhill diagonally
toward the lower left corner, steeply downhill at the highest densities
toward low values of $\log~\Phi_H$, and steeply downhill toward the
upper left corner of each panel. This behavior is explained as
follows.  The ionization parameter increases toward the upper left of
each panel, and the gas becomes highly ionized so that the diffuse
emission becomes a combination of weak free-free and electron
scattering ($\tau_{es}(max) \la 0.07$ for the cloud column density
considered here). Toward the lower left corner, the density diminishes
along constant ionization parameter, and emission lines at diminishing
optical depth cool the gas more efficiently. Toward the lower right
corner, the ionization parameter becomes very small and the electron
temperature drops significantly, squelching the exponential nature of
the free-bound emission. Toward the upper right the lines become
thermalized at large optical depth, and continuum processes (especially
free-bound) cool the gas more efficiently. While the diffuse continuum
emissivity increases toward high gas densities, that which is emitted
at ever shorter wavelengths from a bound-free continuum edge becomes
ever more sensitive to the electron temperature due to the Boltzmann
factor. The temperature is sensitive to the ionization parameter, and
so the emissivity becomes ever more dependent upon the ionizing photon
flux as well. Compare in Figure~1 the contour locations and
orientations of the $\lambda$1793 band to those of the $\lambda$3644
Balmer jump. Optical depth effects play a similar though much more
minor role in governing the diffuse continuum emission efficiencies in
the density -- flux plane (Figure~1). For example, the optical depth
within the Balmer continuum is always larger nearer to the bandhead at
any value of the incident photon flux, and the optical depth increases
with increasing incident photon flux due to increased excited state
populations of hydrogen and larger emission depths. This has the effect
of skewing the emission near the bandhead to larger radii (smaller
incident photon fluxes), as compared to those wavelengths lying ever
shortward of the bandhead.


The band at $\lambda$1218 lies near the peak of the Rayleigh scattering
albedo feature whose strength and wavelength width are in proportion to
the neutral hydrogen column density (see Korista \& Ferland 1998). This
feature generally dominates the diffuse continuum at this wavelength.
This band's contours of intensity relative to the incident continuum
are essentially flat with a value of $\approx$~1 (lower-right half of
the density -- flux plane), since the effective albedo due mainly to
Rayleigh scattering is roughly 1 for neutral hydrogen column densities
exceeding a few $\times 10^{20}$~cm$^{-2}$. Also the contrast of the
Rayleigh scattering feature above the ``background'' (primarily Thomson
scattering) albedo changes in near inverse proportion to the Thomson
scattering albedo, that is in turn proportional to $U_H$ for a fixed
total column density cloud. At sufficiently large ionization
parameters, the hydrogen ionization front emerges out the back ends of
the clouds and the Rayleigh scattering feature disappears rapidly
(upper-left half of the density -- flux plane).

In summary, the diffuse continuum $\lambda\/F_{\lambda}(d)$ for
wavelengths longward of Ly$\alpha$ is near a minimum at the
$\lambda$1458 band, while it reaches its maximum at the Balmer jump for
gas densities exceeding $10^{10}$~cm$^{-3}$, and at the Rayleigh
scattering feature for lower gas densities.

\subsection{The Observed UV-Optical Spectrum of NGC~5548}

We placed the mean {\em HST}/FOS spectrum and the mean optical spectrum
from the 1993 monitoring campaign on the same scale in
$\lambda\/F_{\lambda}$ as best we could. They are plotted in
Figure~2a. The purpose in bringing together these two high quality
spectra was to place modest constraints on the strength of the
predicted diffuse continuum, as presented below. The optical spectrum
is the mean of a subset of the total optical data set that included
only those spectra from the KAST spectrograph of the Lick Observatory
spanning the duration of the {\em HST} portion of the 1993 campaign
(K95). We forced the flux ratio $F_{\lambda}(1350)/F_{\lambda}5100)$ to
match the mean one spanning the {\em HST} portion of the 1993 campaign
(K95).  Finally, the mean {\em HST} and optical spectra were dereddened
with the parameters adopted from Paper~I. Note that the {\em HST}
aperture ($4\arcsec\/.3 \times\/ 1\arcsec\/.4$) was significantly
smaller than the optical aperture. Thus the additive starlight
contribution in the optical spectrum ($\sim$~37\% at 5100~\AA\/; K95)
will be that much larger relative to whatever weak contribution is
present in the UV. We made no attempt to correct for the starlight
contributions, and starlight likely accounts for the optical spectrum's
slight rise in $\lambda\/F_{\lambda}$ toward the red for wavelengths
longer than 4200~\AA\/, as may be perceived in Figure~2a. Notice the
overall flatness of the spectrum in $\lambda\/F_{\lambda}$. This shape
differs somewhat (it's harder) from the SED we adopted for our
photoionization calculations in Paper~I and used here.

\subsection{Model Predictions of the Diffuse Continuum Intensity}

Adopting the same fit of the cloud distribution function parameters to
match the mean UV line spectrum of the 1993 {\em HST} monitoring
campaign as described in Paper~I, we determined the model BLR's
integrated diffuse continuum emission in 35 wavelength bins, spanning
400~\AA\/ to 1.1~\micron\/. In brief, our model-integrated spectra
included clouds distributed spherically around the continuum source
whose gas densities and distances from the central source of ionizing
photons spanned $8 \leq \log n_H~(\rm{cm^{-3}}) \leq 12$ and 1 -- 140
light days (see Paper~I for details). We showed in Paper~I that this
same model was also reasonably successful in reproducing the observed
UV broad emission line variability characteristics. The lower solid
curve in Figure~2a shows the integrated diffuse continuum emission of
the model from Paper~I, appropriately scaled relative to the observed
UV spectrum. To this end, the sum of the diffuse continuum plus some
model for the underlying continuum (discussed below) was forced to pass
through the data at 1145~\AA\/, and the ratio of the diffuse continuum
to the underlying continuum at 1215~\AA\/ then matched the ratio
(0.284) predicted by the model presented in Paper~I. The diffuse
emission declines rapidly shortward of the Rayleigh scattering
Ly$\alpha$ feature within the exponential tail of the Balmer
recombination continuum; just longward of the Lyman jump, the value
drops to 0.016 in units of Figure~2a. The most impressive feature is
the Balmer continuum. This model predicts
$\lambda\/L_{\lambda}(3646)$/$L$(Ly$\alpha) \approx 2.7$, as integrated
over the full BLR. This is large in comparison to ``classical'' BLR
predictions due to the contributions of high-density gas that is
efficient in producing continua, but not Ly$\alpha$. Averaged over the
full UV-optical spectrum, excluding the high-contrast broad emission
lines, the diffuse continuum's contribution to the total UV/optical
nuclear light is $\sim$~15\%.

The shape of the underlying incident continuum is not well constrained.
However, in order to directly compare the approximate contribution of
the predicted diffuse continuum to the data, we considered two {\em ad
hoc} underlying continua to add to the diffuse spectrum. Our first
choice was an $F_{\nu} \propto\/ \nu\/^{-0.7}$ ($\lambda\/F_{\lambda}
\propto\/ \lambda\/^{-0.3}$) power law spectrum (dotted curve in
Figure~2a). The UV portion of the observed spectrum is nearly flat, and
this choice of power law was the flattest possible without exceeding
the data at the Balmer jump, while when combined with the diffuse
continuum also going through the data at 1145~\AA\/. This also happened
to be the approximate slope of the 0.2~keV -- 200~keV spectrum observed
by {\em BeppoSax} in 1997 (Nicastro et al.\ 2000), although it is
unlikely that these power law continua could be the same. The asterisk
icon in Figure~2a indicates the estimated level of the
starlight-subtracted nuclear continuum at 5100~\AA\/ (Romanishin et
al.\ 1995). Such a flat power law clearly passes well above the
expected level at this wavelength.  The observed flux at 5100~\AA\/
would have to lie some 27\% above its plotted position in Figure~2a, in
order for the nuclear continuum (underlying plus diffuse) to pass
through the asterisk. Even given the uncertainties in the interspectrum
normalization and level of starlight contribution, we consider this
unlikely. The larger optical aperture admitted relatively {\em more
starlight} than did the UV aperture, and our choice of interstellar
extinction correction was conservative (Paper~I). Thus improvements in
the UV-optical intercalibration and reddening correction are more
likely to push the optical spectrum a bit {\em lower} relative to the
UV spectrum, than shown in Figure~2a. Next we chose a $F_{\nu}
\propto\/ \nu\/^{1/3}$ ($\lambda\/F_{\lambda} \propto\/
\lambda\/^{-4/3}$) with an appropriate decaying exponential to flatten
and make more convex the UV spectrum in $\lambda\/F_{\lambda}$ (dashed
curve in Figure~2a). This is the rough equivalent of a classical
accretion disk spectrum. When added to the diffuse continuum spectrum,
the sum passes through the data at 1145~\AA\/ and very near the
expected starlight subtracted flux level at 5100~\AA\/. It also passes
just underneath the bulk of the observed UV spectrum, leaving some room
for overlapping broad emission line wing emission and any possible weak
UV starlight contributions. The mismatch shortward of the Balmer jump
may be accounted for by weak starlight and significant overlapping
broad Fe~II emission. It should be also noted that a spectral feature
may appear in the vicinity of the Balmer jump due to the central
continuum source (e.g., Hubeny et al.\ 2000). The mismatch just
longward of the Balmer jump may be accounted for by overlapping high
order Balmer line series, helium lines, Fe~II emission, and some
starlight. Despite the better match, however, we don't place too much
confidence in this particular shape of the underlying continuum. In
particular, this underlying continuum shape is almost certainly too
soft shortward of $\sim$~1150~\AA\/, and too steep for wavelengths
longer than $\sim$~6000~\AA\/ or so, and the actual nuclear light
probably lies between the dashed and dotted curves for these
wavelengths (Figure~2a).

The point of the above exercise was to gain some appreciation for the
relative contribution of the diffuse continuum to the total (diffuse
$+$ underlying) nuclear continuum, as illustrated in Figure~2b. At
$\lambda$1460 the contribution is about 8\%, whereas at $\lambda$5100
the contribution is about 20\%, and at the Balmer jump it is over 40\%
of the total nuclear {\em continuum} flux. The Rayleigh scattering
feature might account for part of the very broad wings of the
Ly$\alpha$ broad emission line. The emission line profile decomposition
of Ly$\alpha$, the results of which were presented in K95, included the
broadest wings of any of the UV emission lines, part of which may be
due to Rayleigh scattering.  However, additional data shortward of
1136~\AA\/ in the rest frame could prove useful in better constraining
the strength of this Rayleigh scattering feature.

It is also apparent from Figures~2a,2b that the contribution from the
diffuse continuum would flatten (in $\lambda\/F_{\lambda}$) an
intrinsically ``blue'' underlying UV/optical continuum. As we discussed
above, it is clear that the nuclear continuum rises toward the UV (in
$\lambda\/F_{\lambda}$), after removing the starlight contribution from
the spectrum. However, the contribution of the diffuse continuum light
emerging from the BLR to the overall nuclear continuum light tends to
increase toward longer wavelengths. Summing the diffuse continuum
spectrum with the second choice of underlying continuum in Figure~2a
changes the spectral power law slope measured between 1350~\AA\/ and
5100~\AA\/ by 0.12 ($-0.63$ to $-0.51$ for $\lambda\/F_{\lambda}
\propto \lambda\/^{\beta}$). This can be important as more
sophisticated spectral models of the central continuum source are
matched to the observed continua of AGN. In a later section we
demonstrate how variability of the diffuse continuum can induce
variability in the measured UV/optical ``continuum'' slope.

In Figure~2c we show the contribution of reflected continuum light to
the total (reflected plus thermal) diffuse continuous light emerging
from the model BLR. Recently, Korista \& Ferland (1998) discussed the
UV/optical albedo of broad line clouds. Thomson scattering and Rayleigh
scattering, as modified by background continuous opacity sources,
predominate through the UV/optical spectrum as sources of reflection.
Given the total column density of the model clouds
($10^{23}$~cm$^{-2}$), at most $\sim$~7\% of the light incident upon a
cloud can be scattered via electron scattering. The effective albedo
will be diminished by continuous opacities, e.g., near bound-free edges
where the thermal diffuse continuous emission is also strongest, and
lie near their expected values where such opacities are smallest, e.g.,
on the longward side of the Balmer jump where the thermal diffuse
continuous emission is also weak. We note that the model integrated
cloud emission did not include the contributions from nearly
transparent clouds of very high ionization parameter ($U_H \gtrsim
30$); no UV/optical emission lines are emitted by this gas. If these
clouds are present with the same distribution functions as adopted for
the UV/optical emission line clouds, the diffuse continuum light in
Figure~2a could be $\sim$~20\% higher due to electron scattering of the
incident continuum by these highly-ionized clouds, except near the
Balmer jump and the Rayleigh scattering feature.

Finally, while our integrated LOC model did not consider the effects of
anisotropic emission from the broad line clouds (or other geometrical
effects), we plot in Figure~2d the ratio of the inward to total diffuse
continuum flux ratio for a slab, as weighted by the adopted model cloud
distribution functions in Paper~I. Davidson \& Netzer (1979), Ferland
et al.\ (1992), and O'Brien, Goad, \& Gondhalekhar (1994) discussed the
anisotropy of line emission from simple broad line clouds. Any
reflection components will be highly anisotropic (inward/total $\approx
1$), by definition, for simple slab clouds. Figure~2d shows a strong
Rayleigh scattering reflection feature, but it also shows enhanced
inward emission on the {\em long} wavelength side of the Balmer
recombination continuum edge. Our model simulations show that the
integrated ratio of the inward to total {\em thermal} diffuse continuum
emission ratio is 0.52 at 3649~\AA\/ and 0.57 at 3644~\AA\/; the
UV/optical thermal diffuse continua just aren't all that optically
thick and so are emitted nearly isotropically. Thus the primary reason
for the significant emission anisotropies shown in Figure~2d is the
reflection of the incident continuum via Thomson and Rayleigh
scattering. The more anisotropically emitting bands will be generally
more sensitive to the emitter/observer geometry, and we have not
considered these complications. However, to keep things in perspective,
the approximate nature of the line and diffuse continuum transfer
formalisms adopted in spectral simulation codes such as {\sc Cloudy} is
as problematic.

\subsection{Model Predictions of the Diffuse Continuum Reverberation}

In this section we will tabulate the predicted diffuse continuum
emission band -- $\lambda$1350 continuum lags, present their light
curves, and attempt to elucidate the effects of the diffuse continuum
on broad emission line -- continuum and continuum -- continuum
reverberation analyses.

\subsubsection{Incident Continuum -- Diffuse Continuum Lags}

We used the observed $\lambda$1350 continuum light curves from the 1989
{\em IUE} and 1993 {\em IUE/HST} monitoring campaigns of NGC~5548
(Clavel et al.\ 1991; Korista et al.\ 1995) to drive our model to
predict the diffuse continuum emitted by the broad line clouds as a
function of time. In other words we assumed that the true driving
continuum, lying mainly at ionizing energies, varies in lock-step with
the $\lambda$1350 continuum. We weighted the response of the cloud
emission by the local responsivity (see Goad, Gondhalekar, \& O'Brien
1993). Once generated, we subtracted off the small diffuse continuum
emission contribution from the observed light at $\lambda$1350 for each
observation date, and then iterated using this slightly altered
continuum driver. This perturbative approach should be relatively
accurate, since the diffuse continuum contribution at this wavelength
is small. As in Paper~I we did not consider any geometrical effects,
such as the anisotropy of the emission. The resulting light curves are
shown in Figures~3a,3b. Other than the $\lambda$1218 and $\lambda$3644
bands that are scaled down by factors of 1.7 and 2, respectively, the
light curves in Figures~3a,3b are plotted to the same scale. Note that
these light curves are not just scaled versions of one another; they
differ in amplitude and some of the bands' light curves weave in and
out of others. This means that the response of the diffuse continuum
will ripple along in wavelength as a function of time. Of all the
bands, those that lie within the hydrogen Paschen continuum (optical
spectrum) respond the most coherently. Figure~3c shows the model rms
variability of the diffuse continuum normalized by the mean diffuse
light as a function of wavelength for both campaigns. Averaged over the
full UV-optical spectrum, this normalized variability amplitude is
roughly two-thirds that of the driving continuum (as represented by the
continuum at $\lambda$1350). As a function of wavelength, it is in
rough inverse proportion to the luminosity-weighted radius of the
emission (larger regions cannot respond as coherently as smaller
ones).  Notice, too, that the strongest diffuse continuum variations
are expected (1) underneath the \ion{C}{4} $\lambda$1549 emission line,
(2) just longward of the Balmer jump, and (3) shortward of
$\sim$~1100~\AA\/. However, Figure~2b shows that the diffuse continuum
is also weak within each of these spectral regions.

Next, we computed the local responsivity-weighted lags for the diffuse
continuum bands relative to the corrected $\lambda$1350 continuum;
these are plotted as a function of wavelength in Figure~4, and
tabulated in Table~1 for those bands plotted in Figure~1. These lags
are determined from the cross-correlation functions computed over the
entire duration of each of these two UV campaigns (White \& Peterson
1994). Note that the lags differ for the two campaigns --- this is due
to the differences in the continuum variability characteristics of each
of the two campaigns.  Figure~4 and Table~1 indicate that one should
expect lags characteristic of the inner broad emission line region,
overlapping with the emission line lags of Ly$\alpha$,
\ion{C}{4}~$\lambda$1549, \ion{Si}{4}~$\lambda$1400, and for spectral
bands shortward of $\sim 1070$~\AA\/ --- more characteristic of
\ion{N}{5}~$\lambda$1240 and \ion{He}{2}~$\lambda$1640. The Rayleigh
scattering feature, effectively due to the extreme damping wings of the
Ly$\alpha$ emission transition, will have lags relative to the incident
continuum that mimic that of the broad emission line of Ly$\alpha$
itself. Like the Ly$\alpha$ emission line, the Rayleigh scattering
feature is expected to be emitted highly anisotropically, though our
model does not take this geometrical complexity into account.

Maoz et al.\ (1993) found an overall lag of $\sim$~10 days between
diffuse emission components in the ``small blue bump'' (after
subtracting off an estimate to the underlying continuum) and the
$\lambda$1350 continuum during the 1989 campaign. This is similar to
our model's predictions given in Table~1 and Figure~4, with an
important caveat being that their derived diffuse emission light curves
included the pseudo-continuum emission of Fe~II in addition to the
Balmer continuum. The last column of Table~1 in Maoz et al., labeled
``Opt1,'' lists their measurements of the diffuse light flux between
3245~\AA\/ and 3635~\AA\/ over the duration of the 1989 campaign,
temporally sparse as they were. They considered this band to contain
primarily Balmer continuum. The mean of their measurements is
approximately $2.8 \times\/ 10^{-12}$ ergs~s$^{-1}$~cm$^{-2}$
(rest-frame). If we divide this wide-band flux by the width of the band
(390~\AA\/) and then multiply by the central wavelength of the band
(3440~\AA\/), we find an approximate mean value of the diffuse flux
$\lambda\/F_{\lambda}(3440) \approx\/ 2.9 \times\/ 10^{-11}$
ergs~s$^{-1}$~cm$^{-2}$ (corrected for reddening, rest frame). This is
very nearly the model's time-averaged value for
$\lambda\/F_{\lambda}(3644) = 2.7 \times\/ 10^{-11}$
ergs~s$^{-1}$~cm$^{-2}$ for the 1989 campaign; see bottom panel of
Figure~3a. In addition the variability amplitude for this model diffuse
continuum band is 1.7, in good agreement with the value measured by
Maoz et al.\ for their ``Opt1'' component (1.6). The model developed in
Paper~I was not constrained by any information pertaining to the Balmer
continuum.  Nevertheless, the model does remarkably well in predicting
the Balmer continuum's overall strength and general variability
characteristics.

Summarizing: wavelength dependencies in variability amplitude and lag
for the diffuse continua are to be expected. The exponential nature of
the bound-free emissivity sensitivity to temperature sets the run of
the luminosity-weighted radius vs.\ wavelength (at wavelengths where
the thermal continua dominate). This means that the lag,
$\tau\/(\lambda)$, of the diffuse continuum relative to the ionizing
continuum is a generally increasing function of wavelength across the
UV-optical spectrum. {\em This general result is not all that sensitive
to the choice of model of the broad line region, but is a simple result
of the gas thermodynamics}.

\subsubsection{Inter-Band Continuum Lags}

What effect does a variable diffuse continuum have upon the
measurements of the so-called ``continuum'' windows? To investigate
this, one can simply assume that the underlying nuclear continuum
varies coherently with wavelength, and then dilute the delay signal of
the diffuse continuum light (Figure~4) by multiplying this by the mean
fraction of the diffuse light to the underlying nuclear continuum light
at each wavelength. The results for both assumptions of the underlying
continuum are shown in Figure~5a,5b. Here it is shown that the
$\lambda$1460 continuum band lies near the minimum in the curves shown,
with a lag of 0.3--0.5 days relative to an unpolluted underlying
continuum. This lag is also typical for the spectral region between
1350~\AA\/ and 1600~\AA\/. However, at 5100~\AA\/ the measured
continuum may lag behind the ionizing continuum by 1--2 days. Thus the
characteristic lag of the broad H$\beta$ emission line, whose light
curve is generally cross-correlated with the continuum band at
5100~\AA\/, may actually be a bit longer relative to the ionizing
continuum than has been reported. Comparison of these UV and optical
continuum band lags leads to an UV-optical inter-band ($\lambda$1350 --
$\lambda$5100) continuum lag of approximately 0.6--1.7 days, again
assuming that the underlying UV-optical continuum varies coherently.
Larger (smaller) UV-optical inter-band lags would be expected if the
underlying UV-optical continuum exhibited a lag relative to the
ionizing continuum proportional (inversely proportional) to wavelength
(see Wanders et al.\ 1997; Collier et al.\ 1999). K95 found a possible
lag between $\lambda$1350 and $\lambda$5100 of 0.7--1.2 days (the
latter lagging behind the former) in NGC~5548, with significant
uncertainty due to the 1 day time-sampling of the campaign. Monte Carlo
simulations indicated a 1.2 day upper limit (90\% confidence level) to
any inter-band continuum lag present in the data. The results presented
here suggest that such a lag may be real and due primarily to diffuse
continuum emission from the BLR. They would also suggest that this
diffuse continuum is at least partially responsible for the systematic
changes in the amplitude and sharpness within the UV continuum band
light curves presented in K95.  It should also be noted that the
typically-chosen continuum windows are likely contaminated with
emission from weak lines and overlapping wings of strong lines at
levels that could rival the diffuse continuum in some of the weaker
diffuse continuum bands.

Hard evidence of a wavelength-dependent UV/optical interband continuum
lag was found in NGC~7469 by Wanders et al.\ (1997), and confirmed by
Collier et al.\ (1999) and Kriss et al.\ (2000). The relationship was
found to be consistent with that expected by the reprocessing of X-rays
illuminating an accretion disk: $\tau\/(\lambda\/) \propto
\lambda\/^{4/3}$ (see also Nandra et al.\ 2000 for a description of
X-ray/UV variability). However, note the similarity of this
relationship, also plotted in Figure~5, to the model
$\tau\/(\lambda\/)$ relation that considers the contribution of a
variable diffuse continuum from the BLR to a coherently varying
underlying continuum. These results bring into question the
interpretation of the origin of the wavelength-dependent interband
continuum lags. Indeed, Figure~14 in Kriss et al.\ (2000) shows that
their longest wavelength UV continuum bands' lags are systematically
too high compared to the illuminated disk model, in just the way
expected if there were significant contamination from diffuse Balmer
continuum from the BLR. The uncertainties in the measured continuum
band lags will need to be reduced substantially to determine whether or
not the accretion disk reprocessing scenario is the correct one, and/or
better constrain the diffuse continuum emission in AGN. For the time
being it would be prudent not to ignore the contribution of the diffuse
continuum from the same gas that emits the lines to the variable
spectra of AGN.

\subsubsection{Bluer When Brighter}

First noticed in NGC~5548 by Wamsteker et al.\ (1990), and again by
Maoz et al.\ (1993) and Korista et al.\ (1995), the UV/optical
continuum, as measured between $\lambda$1350 -- $\lambda$5100, becomes
bluer as it becomes brighter. This effect has also been observed in the
behavior of other Seyfert nuclei variability, though apparently not in
Fairall~9 (see Santos-Lle\'{o} et al.\ 1997). The nature of this effect
has lain in some doubt due to known contamination from non-variable
sources of light, such as starlight that is important primarily at the
longer wavelengths. However, Romano \& Peterson (1998) find this effect
to remain even after subtracting a range of estimates for starlight
contribution at 5100~\AA\/, at least in the case of NGC~5548. For the
starlight contribution determined observationally by Romanishin et
al.\ (1995), and adopted here, Romano \& Peterson find the relation
$\alpha_{1350/5100} = -6.09 \times 10^{-3} F_{\lambda\/1350} + 0.800$,
with $F_{\nu} \propto \nu^{-\alpha}$ defining the logarithmic slope and
$F_{\lambda\/1350}$ in units of
$10^{-15}$~ergs~cm$^{-2}$~s$^{-1}$~\AA\/$^{-1}$. Their relationship did
not take reddening into account, but this affects only the intercept of
their relation. If intrinsic to the nuclear light, the nature of this
relationship is telling us something important about the central
continuum source. A continuum that becomes bluer when brighter may be
indicative of X-ray heating of an accretion disk, for example.

Figure~6a shows the ratio of the flux ($\lambda\/F_{\lambda\/}$) at
$\lambda$1350 to that at $\lambda$5100, after adding the model diffuse
continuum light for each date of the 1989 {\em IUE} campaign to the
underlying continuum at each of the two wavelength bands. We used the
second assumption for the underlying continuum in Figure~2a as a model
for the underlying UV/optical continuum. The underlying UV/optical
continuum's value at 1350~\AA\/ was very nearly the measured value
(corrected by a small amount of diffuse light), and as pointed out
above, its value at 5100~\AA\/ added to the diffuse light there is
nearly equal to the starlight subtracted flux. (Whether or not we
corrected the $\lambda$1350 light curve for a small amount of diffuse
continuum emission had very little impact on the results presented
here.) The model underlying continuum's shape was held fixed, and had a
logarithmic slope of $\alpha_{1350/5100} = 0.363$ ($F_{\nu} \propto
\nu^{-\alpha}$). Figure~6a shows that the predicted flux ratio
fluctuates by $\pm$5\% about a mean of 1.98, and is correlated with the
underlying continuum at $\lambda$1350 at the 87\% level with the ratio
leading the underlying continuum by a couple of days (see also the top
panel of Figure~3a). In other words, the effective nuclear light
becomes bluer when it becomes brighter due to the reverberation of the
diffuse continuum within the BLR. The same effect was seen in similar
simulations of the 1993 {\em HST} campaign, with slightly smaller
fluctuations about a mean ratio of 1.91, again with the continuum flux
ratio leading slightly the underlying continuum.  Figure~6b shows the
effective logarithmic slope, $\alpha_{1350/5100}$, as a function of the
flux of the underlying continuum at $\lambda$1350 for both monitoring
campaigns. The effective slope was determined from the flux ratio
above, and so considers the sum of the underlying plus diffuse
continuum components. We adopted the same format as in Figure~3 of
Romano \& Peterson to ease comparison. This slope is highly correlated
with the underlying $\lambda$1350 continuum in both campaigns, though
with differing relationships --- likely due to the differences in the
continuum variability characteristics of each of the two campaigns.
Note, too, that the effective logarithmic slope may deviate from that
of the underlying continuum (0.363) by up to 0.2 (with
$\alpha_{eff}\rm{(max)} \sim\/ 0.56$), with the total UV/optical
continuum light being significantly softer than the underlying nuclear
continuum.  Greatest deviations between the underlying and effective
UV/optical slopes are expected when the underlying continuum is {\em
low} (assuming that the ionizing and underlying continua are
well-correlated). Figure~3c shows that the diffuse continuum at
1350~\AA\/ is somewhat more variable and Figure~4 shows it to have a
smaller lag than the diffuse continuum at 5100~\AA\/. The diffuse
continuum is less variable than the underlying continuum at all
wavelengths. Taken together these account for the results presented in
Figures~6a,6b. However, comparing the slopes of the lines fitted to the
points in Figure~6b to the slope of the relation found by Romano \&
Peterson (above), one sees that the reverberation of the diffuse
continuum emitted by the BLR can account for only $\sim$~1/3 of the
observed effect, meaning that the underlying continuum likely does
become bluer as it becomes brighter.  What we have demonstrated here is
that not all of this effect should be automatically attributed to the
underlying continuum.

\section{DISCUSSION \& SUMMARY}


\subsection{The Rayleigh and Electron Scattering Features}

The shortest wavelength continuum band reported by K95, $\lambda$1145,
lies within the blue wing of the predicted Ly$\alpha$ Rayleigh
scattering feature in Figure~2a. In the spectrum from any single cloud
the width of this feature is proportional to the neutral column density
(Korista \& Ferland 1998), and unlike all other features in the diffuse
continuum spectrum, it is observed only in reflection. While the light
curve data in this band were noisy, K95 reported that this band had the
largest amplitude and sharpest variations, though neither
characteristic was much different from those in the $\lambda$1350
band. Nevertheless, based upon the observed characteristics of the
light curves one might expect less contamination of diffuse continuum
emission from the BLR within the $\lambda$1145 band than elsewhere, in
possible contradiction to the model results. Extending the observed
spectral coverage to shorter wavelengths should help constrain the
model predictions of this scattering feature. It is also clear that
this feature contributes to the measured Ly$\alpha$ broad emission line
flux and has similar reverberation properties.

The effects of electron scattering on the diffuse continuum spectrum
are fairly minimal in these simulations given our choice of constant
column density clouds with $\log\/ N_H = 23$ (see Figure~2c). As
mentioned above, however, most of the anisotropy in the diffuse
emission from the clouds (Figure~2d) is expected to arise from the
electron scattering component (well outside the Rayleigh scattering
feature). For wavelengths well longward of the Ly$\alpha$ Rayleigh
scattering feature ($\lambda \gtrsim\/ 1800$~\AA\/) the reflected-light
component of the diffuse continuum, mainly Thomson scattering of the
incident continuum, has a luminosity-weighted radius that is roughly
30\% that of the Ly$\alpha$ emission line and the peak of the Rayleigh
scattering feature (all of which are predicted to be heavily inwardly
beamed). This is expected in models that geometrically weight the
contributions of light from clouds nearer to the continuum source ---
the stronger electron-scattering mirrors are those clouds with larger
ionized hydrogen column densities. The expected lag of this
reflected-light component is very nearly independent of wavelength
across the UV/optical spectrum, in contrast to that of the thermal
diffuse light for which the lag generally increases with wavelength,
and whose emission is expected to be much less anisotropic. Thus larger
contributions from electron scattering of the incident continuum off
the broad line clouds (1) will result in a greyer response; (2) the
anisotropic nature of the electron-scattered light may increase the
amplitude of the response by reducing the range in time-delays over
which we see the diffuse light; (3) the anisotropic nature of the
electron-scattered light will offset the electron-scattered light's
tendencies to have smaller luminosity-weighted radii than the thermal
diffuse continuum.

\subsection{The Lyman and Balmer Continua}

The model predicts a Lyman jump in emission with an intensity similar
to that of the peak of the Ly$\alpha$ Rayleigh scattering feature in
Figure~2a, amounting to a $\sim$~25\% jump in the total nuclear light
near 912~\AA\/, assuming the  underlying continuum is featureless in
the vicinity of the jump. After considering spectral broadening due to
bulk gas motions, this jump might only be $\lesssim$~20\% --- but still
easily observable. Our model
$\lambda\/L_{\lambda}(912)$/$L$(Ly$\alpha)$ is roughly 2.4, large in
comparison to ``classical'' BLR predictions due to the contributions of
high-density gas ($n_H > 10^{10}$~cm$^{-3}$) that is efficient in
producing continua, but not Ly$\alpha$. It is important to note that
the prediction of a Lyman feature in emission is generic to virtually
all spectral simulations of the broad-line emitting gas. However, while
the Balmer continuum is observed within the little blue bump, the Lyman
continuum is {\em never} observed. In fact the quasar spectra in this
region are remarkably featureless (Koratkar \& Blaes 1999), with only
hints of a small {\em depression} near 912~\AA\/ in some objects (e.g.,
Kriss et al.\ 1995). This has been a long-standing puzzle for models of
both broad line clouds and accretion disks, though the more recent
considerations of Comptonization (Kriss et al.) and non-LTE effects
(Hubeny et al.\ 2000) generally produce weaker absorption features in
the vicinity of the Lyman limit than the earlier models of accretion
disk spectra. As averaged over our BLR model, some 90\% of the diffuse
continuum at the Lyman jump is beamed inward, and perhaps this plays
some role in hiding it from view. It is also conceivable that a
conspiracy of sorts exists whereby the accretion disk feature in
absorption near 912~\AA\/ counteracts the feature expected in emission
from the broad line gas, as suggested by Carswell \& Ferland (1988).
But these are rather unsatisfying scenarios, and we must admit we do
not have a good understanding of the Lyman continuum spectral feature.

Since the Balmer continuum is observed in emission, an understanding of
its origin would constrain models of broad emission line gas as well as
those of accretion disks --- the Balmer continuum could appear in
emission in a chromosphere of an accretion disk. If the Balmer
continuum is emitted primarily in the BLR, a significant lag ($\sim$~1
week for NGC~5548) is expected between flux variations in the emission
near the jump and those of the underlying continuum. If, on the other
hand, the Balmer continuum is emitted in the chromosphere of an
illuminated accretion disk, the lags would be expected to be much
smaller. In fact in this case, the wavelength-dependent interband
continuum lag would deviate from the simple $\lambda\/^{4/3}$
relationship, and behave in a way similar to that presented in
Figure~5a except that the lags should be much smaller across the Balmer
continuum. The results of the analysis of Maoz et al.\ (1993) would
appear to favor an origin in the BLR, but this must be confirmed with
follow-up observations of NGC~5548 and other AGN. Weak Balmer and
Paschen continua from broad line gas is a signature of either the
predominance of low density gas ($n_H \ll 10^{12}$~cm$^{-3}$), or the
presence of a mechanism that broadens the local line profiles
significantly above their thermal values. Local line broadening reduces
the line optical depths, effectively increasing the cooling efficiency
of the lines. Bottorff et al.\ (2000) explored the effect of local line
broadening to quasar spectra. Simulations from that work (Bottorff
2000) and some of our own show that by increasing the local line width
from thermal to 1000~$\rm{km~s^{-1}}$, the model luminosity of the
Balmer jump ($\lambda\/L_{\lambda}(3646)$) falls by a factor of 2.5,
and its ratio with respect to the luminosity of Ly$\alpha$ falls by a
factor of 4.8.  In addition, the strength of the Balmer jump relative
to Ly$\alpha$ and the strengths of the diffuse recombination continua
in general are very sensitive to the presence and importance of high
density gas ($n_H > 10^{10}$~cm$^{-3}$), especially if the local line
widths are thermal.  The presence of super-solar metallicity gas would
also diminish the thermal diffuse continua, due to the accompanying
decreases in electron temperatures, though these metallicities are not
expected to occur in Seyfert~1 nuclei. Again, we emphasize that these
trends are general and not dependent upon our particular model of the
BLR of NGC~5548. It is clear that the under-explored Balmer continuum
should be mined for its important physical constraints.

\subsection{Interband Continuum Variations in Other AGN}

Short, but intense, multiwavelength monitoring campaigns have been
mounted to detect correlated variability amongst the various energy
bands (X-ray, UV, optical) emitted by Seyfert~1 nuclei. Edelson et
al.\ (1996) reported good correlations between X-ray, UV, and optical
continuum variations during the 9-day monitoring campaign of NGC~4151,
but could not measure any lags within or between any of these bands,
and set a limit of $\lesssim$~1 day lag between $\lambda$1275 and
$\lambda$5125. In the 3-day campaign of NGC~3516 Edelson et al.\ (2000)
reported correlated hard and soft X-ray variations and correlated
optical variations, but the X-ray and optical bands showed very
different variability behavior. They set a limit of $\lesssim$~0.15 day
lag between the $\lambda$3590 and $\lambda$5510 bands, and suggested
that their findings represent serious problems for ``reprocessing''
models in which the X-ray source heats the accretion disk. It is
unlikely that either of these experiments were long enough for the
continuum variations to sufficiently sample the BLRs in order to detect
variations from the diffuse continuum. This is in contrast to the high
temporal resolution campaign of NGC~7469 that spanned 50 days. As
discussed above, significant lags were found across its UV and optical
continuum bands, and were found to be consistent with a simple model of
an X-ray illuminated accretion disk. However, the results presented
here suggest that this conclusion may be premature, as the variable
diffuse continuum from the BLR may broadly mimic the signature of an
X-ray heated accretion disk.

\subsection{Summary}

Unless some mechanism is present that acts to suppress the emission of
diffuse thermal continua, that which is emitted in the UV/optical by
broad line clouds will be significant. Its presence in combination with
the light reflected from the broad line clouds has the following
impacts on models for the broad emission line region and the UV-optical
continuum emitter: (1) the lag between the emission line variations and
those of the local ``continuum'' may underestimate the characteristic
sizes of the line emitting regions, (2) the shape of the underlying
nuclear continuum may be misinferred and thus misinterpreted, (3)
interband continuum lags may be misinterpreted, (4) the reverberation
of the diffuse continuum emitted by the BLR may account for perhaps
one-third of the observed effect that the $\lambda$1350 --
$\lambda$5100 continuum becomes bluer as it becomes brighter (even
after accounting for non-variable optical starlight), and (5) as in
other models of the broad emission line gas, a significant yet
unobserved Lyman jump is predicted. The underexplored diffuse
continuum, most easily observable near the Balmer jump, holds clues to
our understanding of both the line and central continuum emitting
regions in AGN. Long-term, high temporal resolution, spectroscopic
monitoring with full UV-optical spectral coverage is required to
understand the nature of UV-optical continuum and its variability in
AGN. This can be accomplished with a dedicated multiwavelength orbital
platform.

\acknowledgements

We are grateful to Gary Ferland for maintaining his freely distributed
code, {\sc Cloudy}, to Brad Peterson for supplying the mean optical
spectrum of NGC~5548, and to Mark Bottorff for supplying some spectral
simulation results. We also acknowledge helpful conversations with Stefan
Collier, and the constructive comments of an anonymous referee.

%
%

%
\clearpage

\begin{figure}
\figurenum{1a}
\caption{Contours of theoretical
$\log[\lambda\/F_{\lambda\/}\rm{(diffuse)} /
\lambda\/F_{\lambda\/1215}\rm{(incident)}]$ for 6 diffuse continuum
bands are shown in the plane of hydrogen gas density and flux of
hydrogen ionizing photons. The total hydrogen column density is
$10^{23}~\rm{cm^{-2}}$, and each point in the grid assumes full source
coverage. The smallest, generally outermost, solid decade contours
correspond to a ratio of 0.1, solid lines are decades, and dotted lines
represent 0.1 decade steps between. The contours generally decrease
monotonically from the peak (solid triangle, with ratio values $\sim$~1
at very high gas densities and photon fluxes) to the ratio = 0.1
contour. The solid star is a reference point marking the old
``standard BLR'' parameters.}
\end{figure}

\begin{figure}
\figurenum{1b}
\caption{Same as Figure~1a for 6 more diffusing continuum bands.}
\end{figure}

\begin{figure}
\figurenum{2a}
\caption{The mean, dereddened UV-optical spectrum of NGC~5548 from the
1993 {\em HST} campaign (top, solid), the appropriately scaled model
diffuse continuum (bottom, solid), and the sum of the latter and two
guesses to the underlying nuclear continuum (middle):
$\lambda\/F_{\lambda} \propto \lambda\/^{-0.3}$ (dotted),
$\lambda\/F_{\lambda} \propto
\lambda\/^{-4/3}e^{-\lambda\/_{cut}/\lambda\/}$ (dashed). The most
prominent features in the diffuse continuum are the Balmer continuum
and the Ly$\alpha$ Rayleigh scattering feature. The Paschen continuum
dominates the optical regime. The asterisk represents the approximate
starlight-subtracted continuum flux level at 5100~\AA\/.}
\end{figure}

\begin{figure}
\figurenum{2b}
\caption{The fractional contribution of the diffuse continuum to the
sum diffuse plus underlying continua for both models of the underlying
continuum. Dashed and dotted symbols refer to the corresponding
underlying continua assumed in Figure~2a.}
\end{figure}

\begin{figure}
\figurenum{2c}
\caption{The fractional contribution of the reflected incident
continuum to the total diffuse light emerging from the model BLR.}
\end{figure}

\begin{figure}
\figurenum{2d}
\caption{The integrated LOC model ratio of the inward to total diffuse
continuum flux emitted by the slab-cloud geometry assumed in the {\sc
Cloudy} computations as a function of wavelength. Values of 0.5
indicate isotropic emission, whereas values of 1 indicate fully
inwardly beamed emission.}
\end{figure}

\begin{figure}
\figurenum{3a}
\caption{Top panel: the measured, rest frame, reddening corrected,
continuum light curve at 1350~\AA\/ from the 1989 {\em IUE} campaign of
NGC~5548, corrected also for the diffuse continuum emission at
1350~\AA\/. Bottom two panels: model diffuse continuum light curves
using the continuum in the top panel as the driver of the BLR response. }
\end{figure}

\begin{figure}
\figurenum{3b}
\caption{Same as Figure~3a but for the 1993 {\em HST} campaign.}
\end{figure}

\begin{figure}
\figurenum{3c}
\caption{The model rms variation normalized by the mean light for the
diffuse continuum as a function of wavelength for the 1989 (bold) and
1993 (thin) campaigns. The rms variation is roughly inversely
proportional to the luminosity-weighted radius of the diffuse continuum
emission.}
\end{figure}

\begin{figure}
\figurenum{4}
\caption{The wavelength-dependent lag of the diffuse continuum relative
to the corrected $\lambda$1350 continuum driver. The bold solid and
dashed lines are the model peak and centroid CCF lags derived using the
corrected $\lambda$1350 continuum of the 1989 {\em IUE} campaign as the
driver. The thin solid and dashed lines are the model peak and centroid
CCF lags derived using the corrected $\lambda$1350 continuum of the
1993 {\em HST/IUE} campaign as the driver.}
\end{figure}

\begin{figure}
\figurenum{5a}
\caption{Same as Figure 4, but with the lags scaled by the ratio of the
diffuse continuum to the underlying continuum. Here it is assumed that
the underlying continuum varies {\em coherently} in wavelength, and it
has the form $\lambda\/F_{\lambda} \propto \lambda\/^{-0.3}$ (see
Figure~2a). Differences in lag should be representative of the lag
measured between the light curves of any two so-called continuum
windows. The extra-bold dash-dotted curve is the function
$\tau\/(\lambda\/) = 3(\lambda\//10^4 \rm{\AA\/})^{4/3}$ from Kriss et
al.\ (2000).}
\end{figure}

\begin{figure}
\figurenum{5b}
\caption{Same as Figure 5a, but assuming an underlying continuum of the
shape $\lambda\/F_{\lambda} \propto
\lambda\/^{-4/3}e^{-\lambda\/_{cut}/\lambda\/}$ (see Figure~2a). The
extra-bold dash-dotted curve is the function $\tau\/(\lambda\/) =
4.5(\lambda\//10^4 \rm{\AA\/})^{4/3}$.}
\end{figure}

\begin{figure}
\figurenum{6a}
\caption{The ratio of the model diffuse plus underlying continua at
1350~\AA\/ to that at 5100~\AA\/ as a function of time, for the 1989
{\em IUE} campaign. This ratio is highly positively correlated with the
flux at 1350~\AA\/, and so the $\lambda$1350 -- $\lambda$5100 continuum
becomes bluer as the underlying continuum becomes brighter.}
\end{figure}

\begin{figure}
\figurenum{6b}
\caption{The logarithmic slope, $\alpha_{1350/5100}$, as a function of
the flux at $\lambda$1350 for both monitoring campaigns (squares: 1989
{\em IUE}; triangles: 1993 {\em HST}). Simple linear fits are also
shown for each. The logarithmic slope of the model underlying continuum
was fixed at $\alpha_{1350/5100} = 0.363$ ($F_{\nu} \propto
\nu^{-\alpha}$).}

\end{figure}

\clearpage

%
\begin{deluxetable}{lcc}
\small
\tablecaption{Model Diffuse Continuum Band Lags. 
\label{tbl-1}}
\tablewidth{0pt}
\tablehead{
\colhead{Wavelength} & \colhead{$\tau_{\eta}(IUE89)$} &
\colhead{$\tau_{\eta}(HST93)$}
}
\startdata
$\lambda$1143 & 5.3--6.1  & 3.5--6.5  \\
$\lambda$1218 & 9.7--10.6 & 6.5--9.4  \\
$\lambda$1356 &  4.9--5.2 & 3.4--6.0 \\
$\lambda$1458 &  4.8--4.9 & 3.4--5.9  \\
$\lambda$1793 &  5.2--5.6 & 3.6--6.5 \\
$\lambda$2071 & 5.7--6.3 & 4.5--7.2  \\
$\lambda$2392 & 6.2--7.6 & 5.4--7.9  \\
$\lambda$3644 & 10.0--11.1 & 7.5--10.2 \\
$\lambda$3649 & 5.4--6.0 & 4.4--6.8  \\
$\lambda$4885 & 6.0--7.0 & 4.6--7.5  \\
$\lambda$6209 & 6.3--7.9 & 5.4--8.0  \\
$\lambda$8205 & 6.7--8.6 & 5.5--8.5  \\
\enddata

\tablecomments{The local responsivity-weighted CCF lags, $\tau_{\eta}$,
are given in units of days. Each pair of lags relative to the driving
continuum ($\lambda$1350) is given as ``CCF peak -- CCF centroid,'' the
latter measured at 50\% of the CCF peak. All lags have been determined
from the full durations of the campaigns.}

\end{deluxetable}


\clearpage
\begin{references}

\reference{} Baldwin, J.A., Korista, K.T., Ferland, G.J., \& Verner,
D.A.\ 1995, \apj, 455, L119

\reference{} Bottorff, M.\ 2000, private communication

\reference{} Bottorff, M., Ferland, G., Baldwin, J., \& Korista,
K.\ 2000, \apj, 542, 644

\reference{} Carswell, R.F., \& Ferland, G.J.\ 1988, \mnras, 235, 1121

\reference{} Clavel, J., et al.\ 1991, \apj, 366, 64

\reference{} Collier, S.\ Horne, K., Wanders, I., \& Peterson,
B.M.\ 1999, \mnras, 302, 24

\reference{} Davidson, K.\ 1976, \apj, 207, 710 

\reference{} Davidson, K., \& Netzer, H.\ 1979, Rep.\ Prog.\ Phys., 51,
715

\reference{} Edelson, R., et al.\ 1996, \apj, 470, 364

\reference{} Edelson, R., et al.\ 2000, \apj, 534, 180

\reference{} Ferland, G.J., Peterson, B.M., Horne, K.D., Welsh, W.F., \&
Nahar, S.N.\ 1992, \apj, 387, 95

\reference{} Ferland, G.J., 1997, HAZY, A Brief Introduction to
{\sc Cloudy} (Univ.\ Kentucky Phys.\ Dept.\ Int.\ Rep.)

\reference{} Ferland, G.J., Korista, K.T., Verner, D.A., Ferguson, J.W.,
Kingdon, J.B., \& Verner, E.M.\ 1998, \pasp, 110, 761


\reference{} Goad, M.R., O'Brien, P.T., \& Gondhalekar, P.M.\ 1993,
\mnras, 263, 149


\reference{} Hubeny, I., Algol, E., Blaes, O., \& Krolik, J.H.\ 2000,
\apj, 533, 710

\reference{} Kaspi, S., \& Netzer H.\ 1999, \apj, 524, 71

\reference{} Koratkar, A., \& Blaes, O.\ 1999, \pasp, 111, 1

\reference{} Korista, K.T., et al.\ 1995, \apjs, 97, 285 (K95)

\reference{} Korista, K., Baldwin, J., Ferland, G., Verner, D.\ 1997,
\apjs, 108, 401

\reference{} Korista, K., \& Ferland, G.\ 1998, \apj, 495, 672

\reference{} Korista, K.T., \& Goad, M.R.\ 2000, \apj, 536, 284

\reference{} Kriss, G.A., et al.\ 1995, \apj, 444, 632

\reference{} Kriss, G.A., Peterson, B.M., Crenshaw, D.M., \& Zheng,
W.\ 2000, \apj, 535, 58

\reference{} Maoz, D., et al.\ 1993, \apj, 404, 576

\reference{} Nandra, K., Le, T., George, I.M., Edelson, R.A.,
Mushotzky, R.F., Peterson, B.M., \& Turner, T.J.\ 2000, \apj, in press

\reference{} O'Brien, P.T., Goad, M.R., \& Gondhalekar, P.M.\ 1994,
\mnras, 268, 845

\reference{} Romanishin, W.\ et al.\ 1995, \apj, 455, 516





\reference{} Rees, M.J., Netzer, H., \& Ferland, G.J.\ 1989, \apj, 347,
640

\reference{} Romano, P., \& Peterson, B.M.\ 1998, \apj, in preparation,
(astro-ph/9806190)

\reference{} Santos-Lle\'{o}, M., et al.\ 1997, \apjs, 112, 271

\reference{} Wamsteker, W., et al.\ 1990, \apj, 354, 446

\reference{} Walter, R., Orr, A., Courvoisier, T.J.-L., Fink, H.H.,
Makino, F., Otani, C., Wamsteker, W.\ 1994, \aap, 285, 119

\reference{} Wanders, I., et al.\ 1997, \apjs, 113, 69

\reference{} White, R.J.\ \& Peterson, B.M., 1994, \pasp, 106, 879

\end{references}
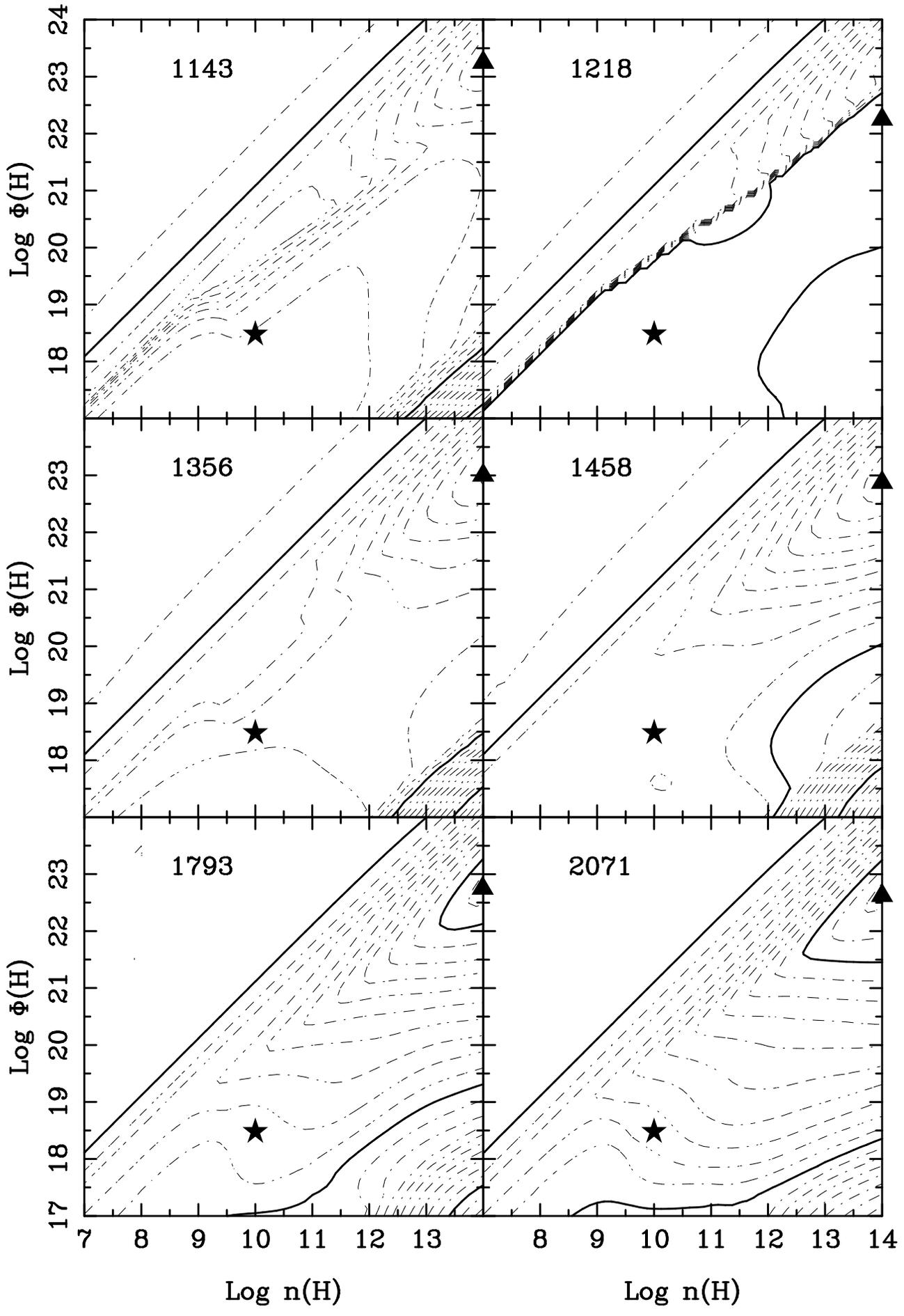
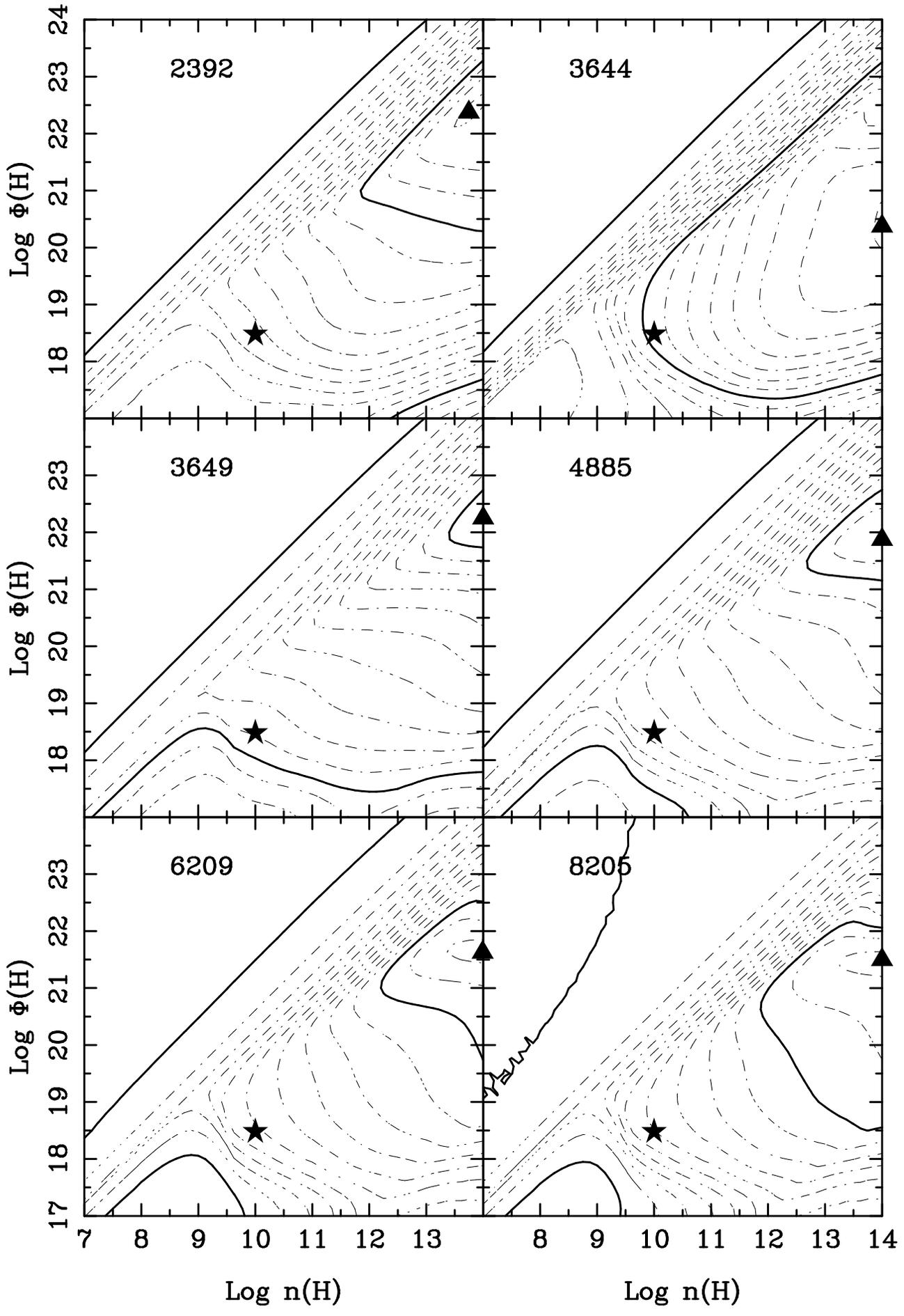
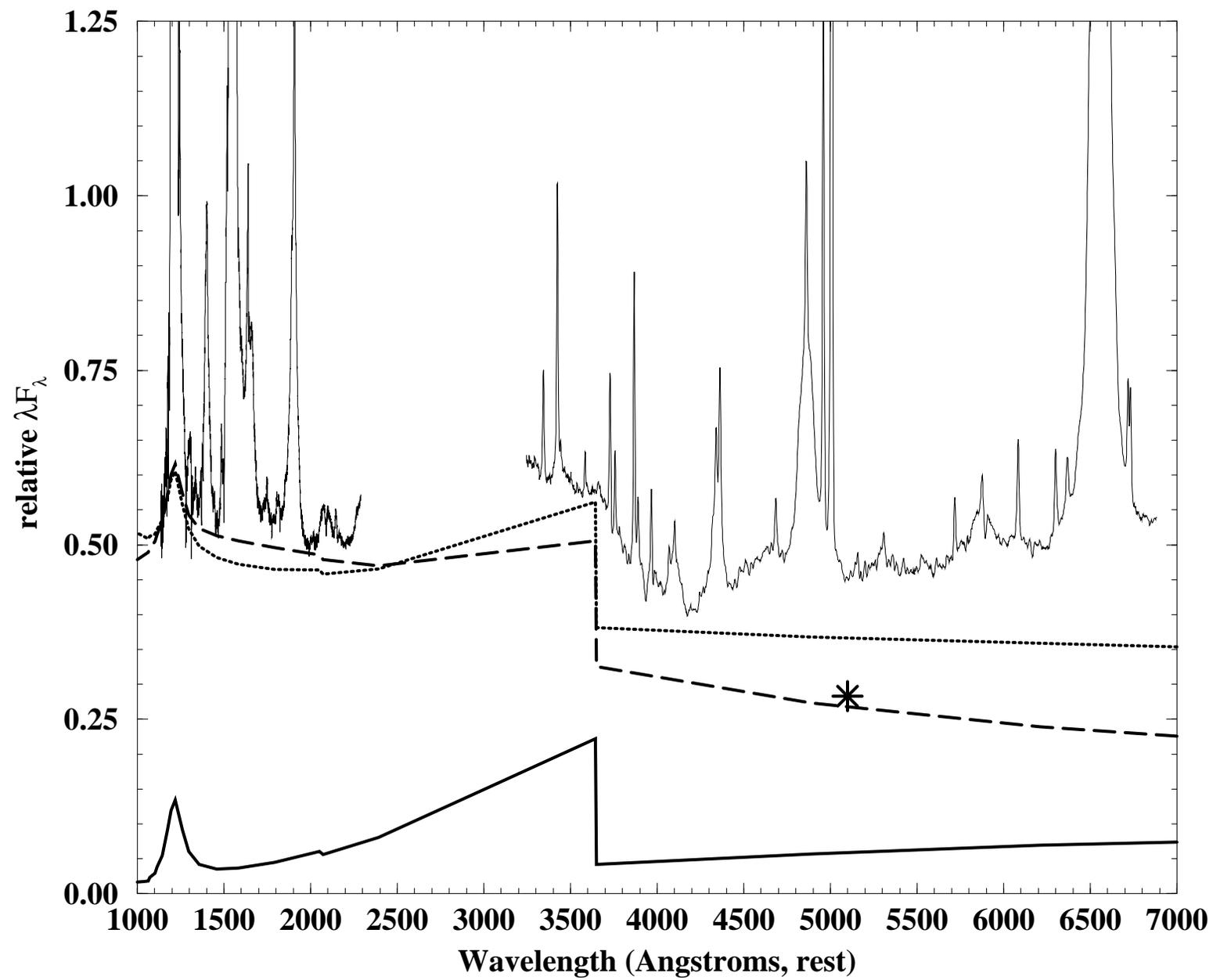
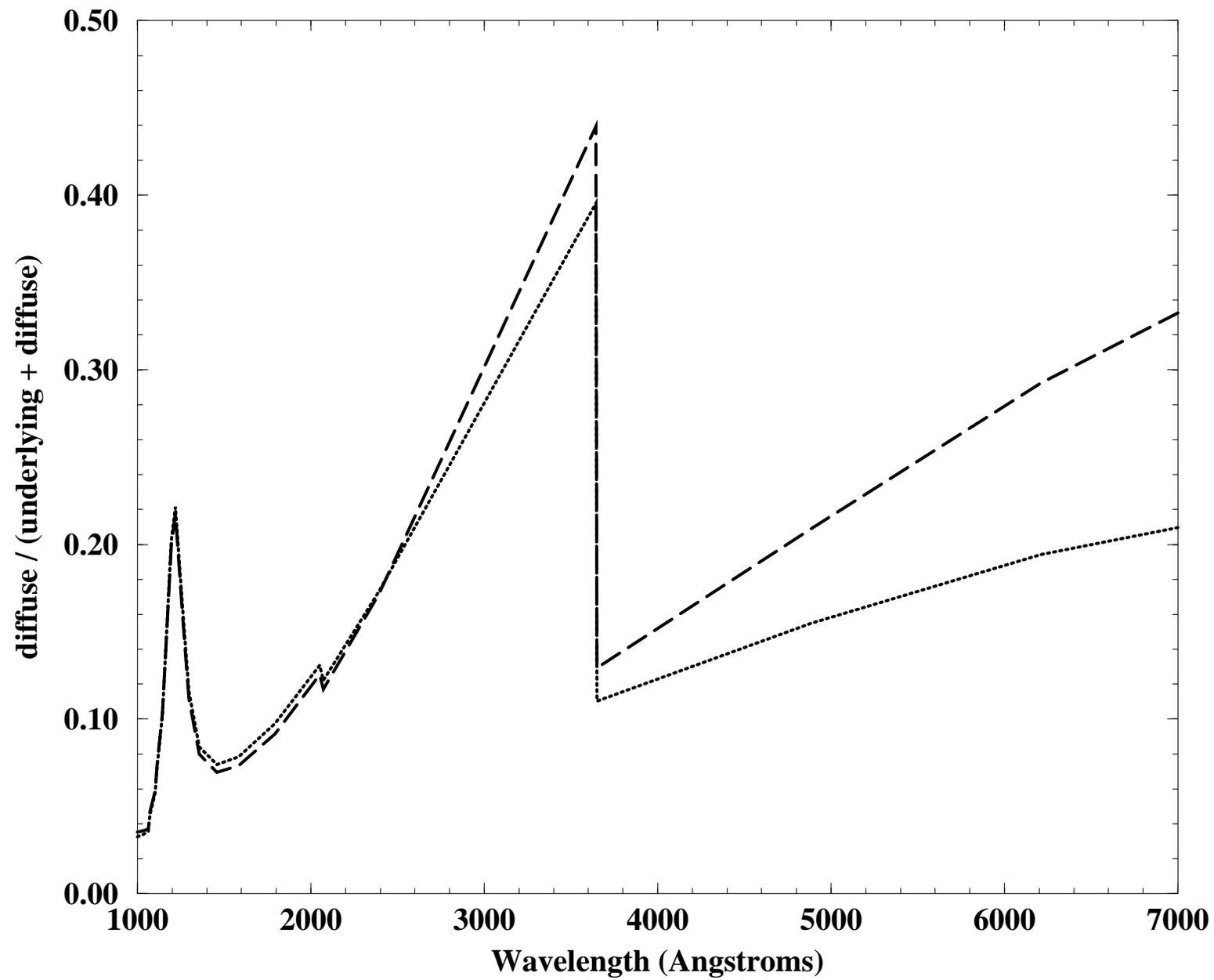
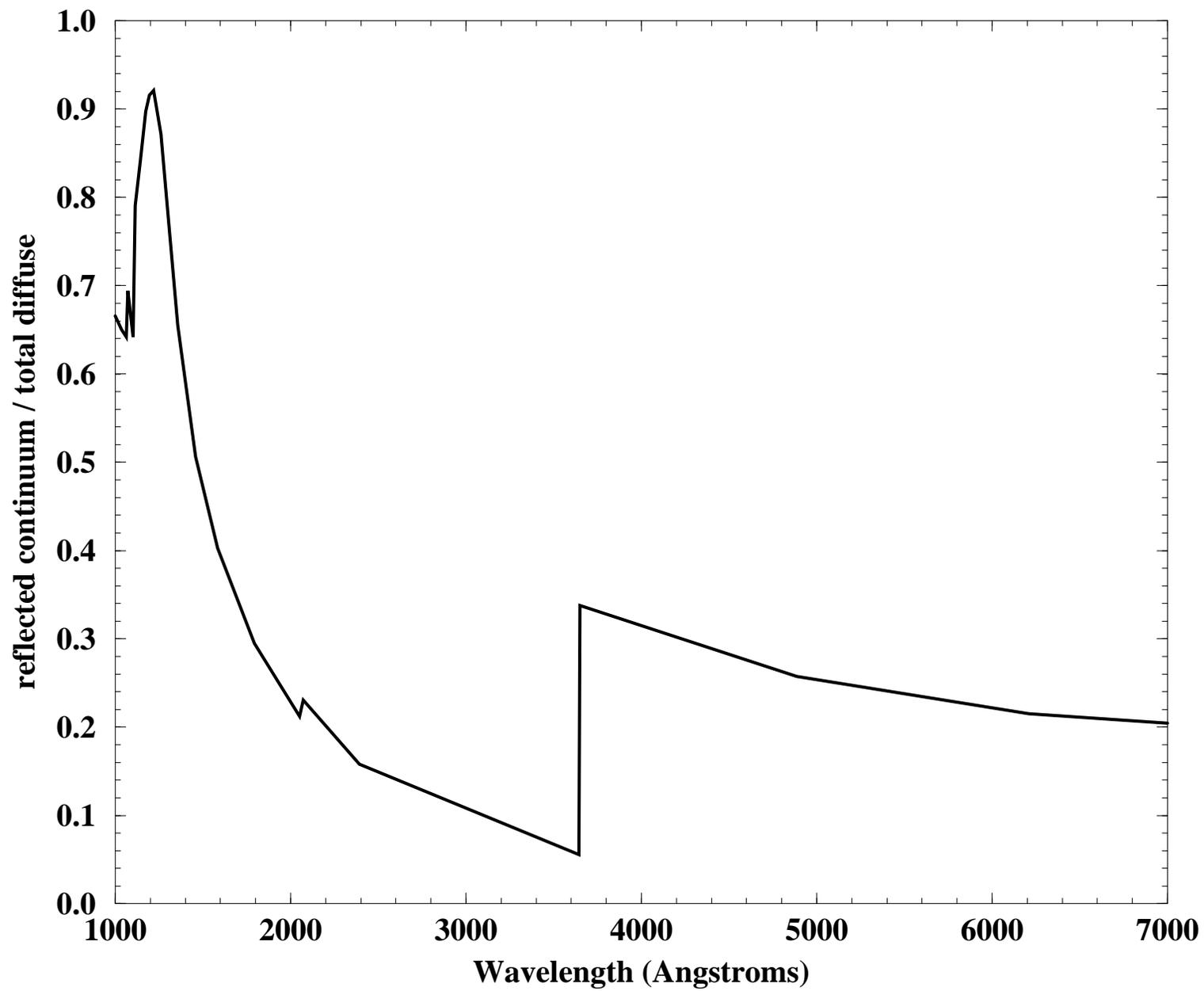
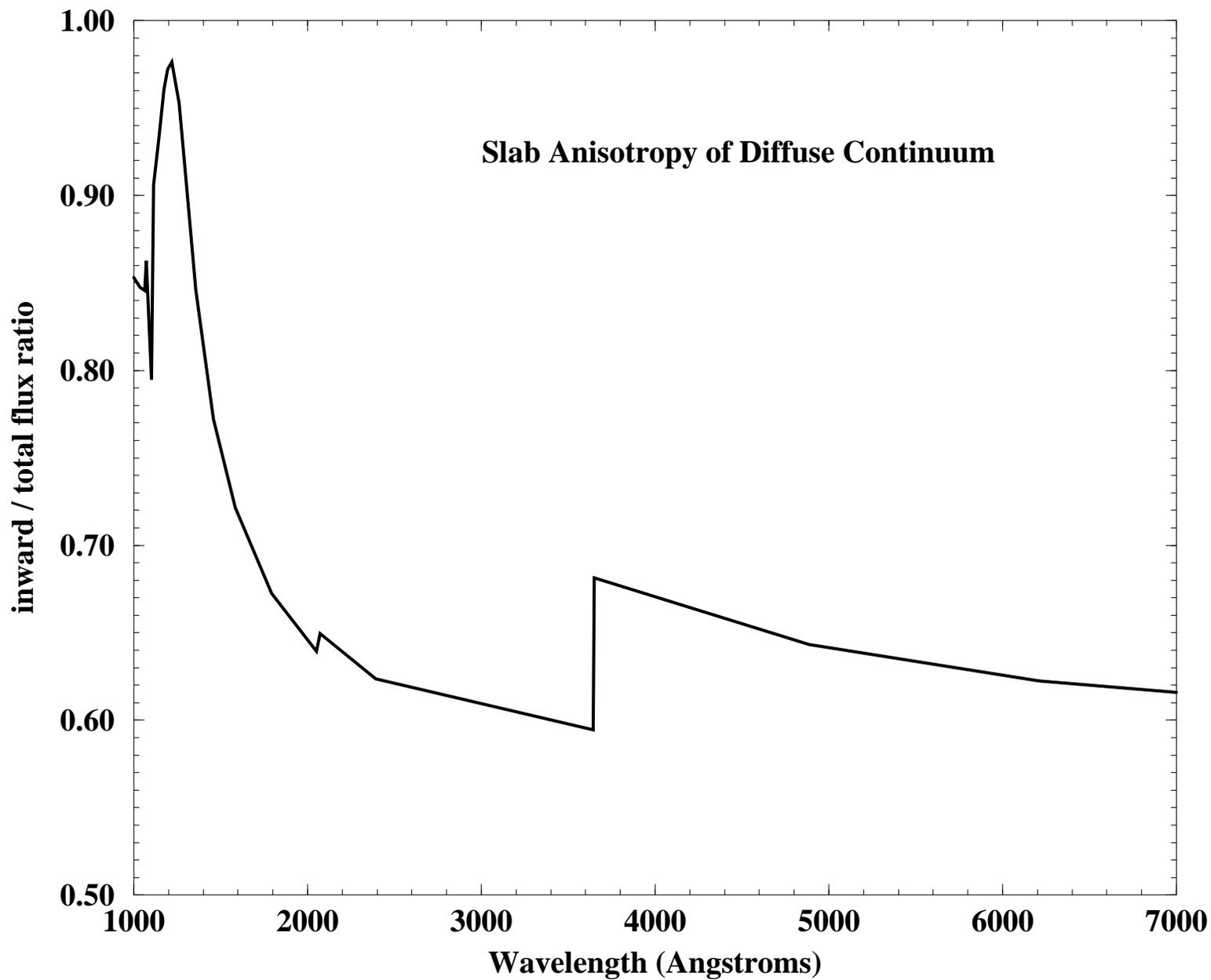
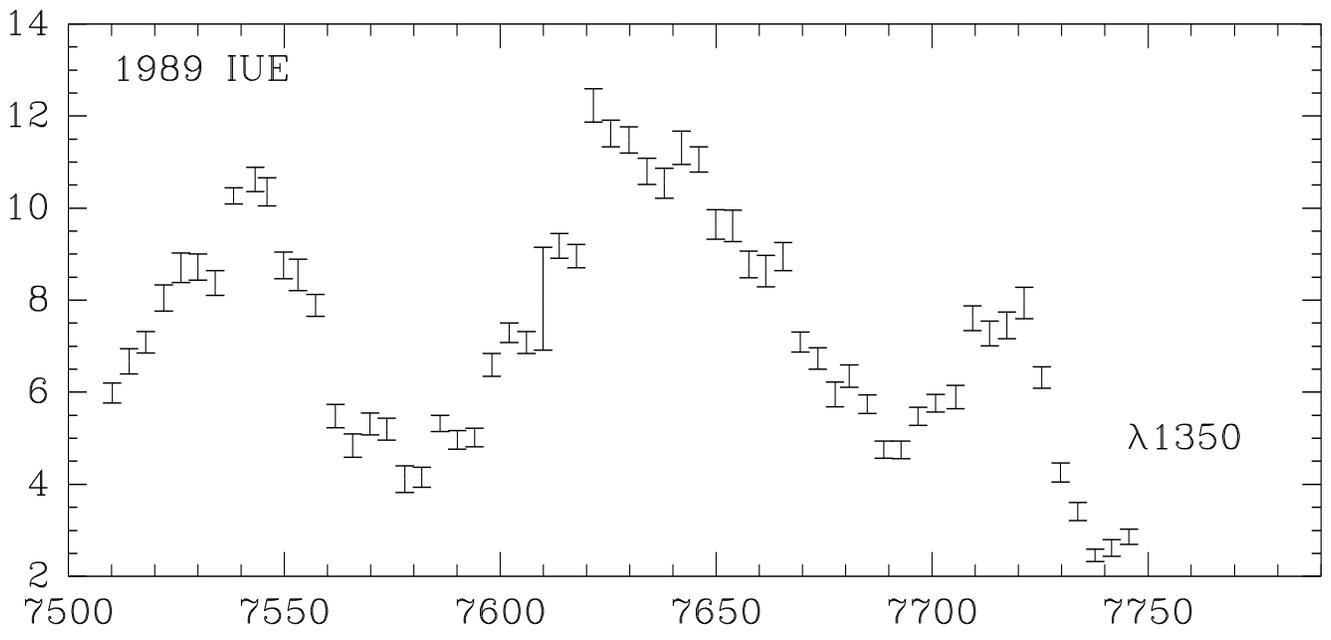
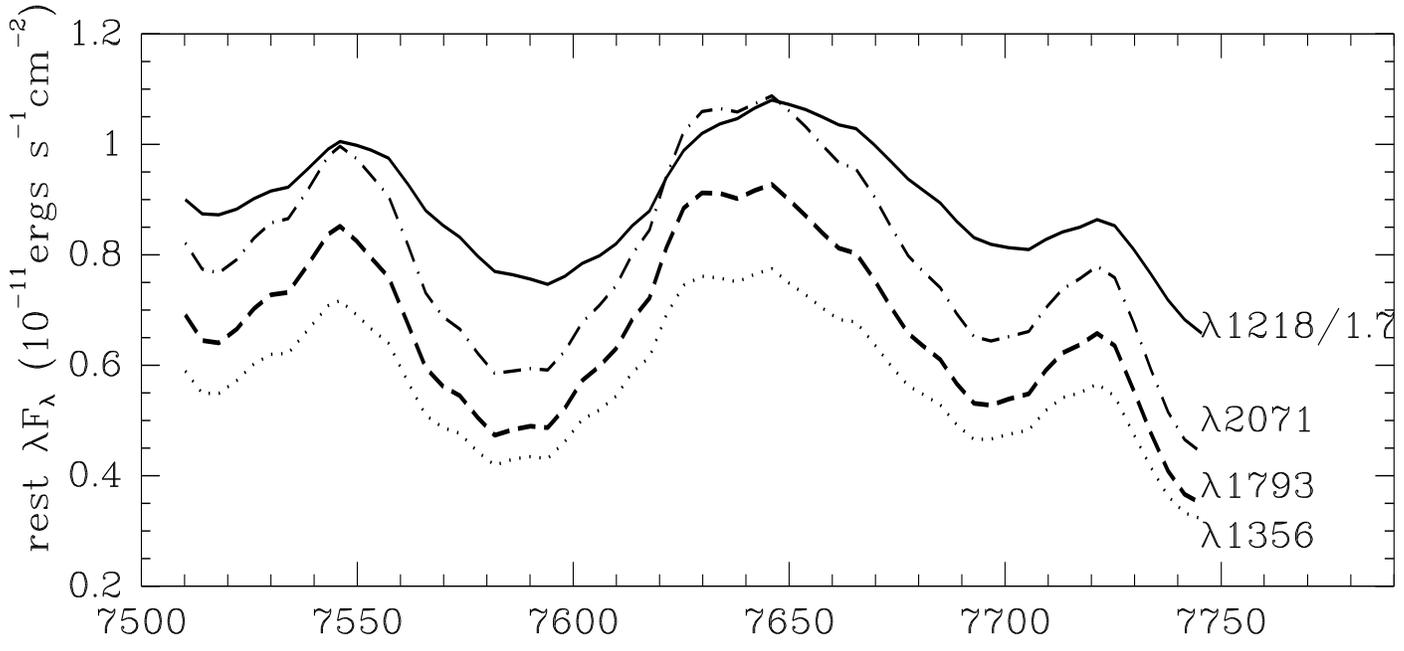
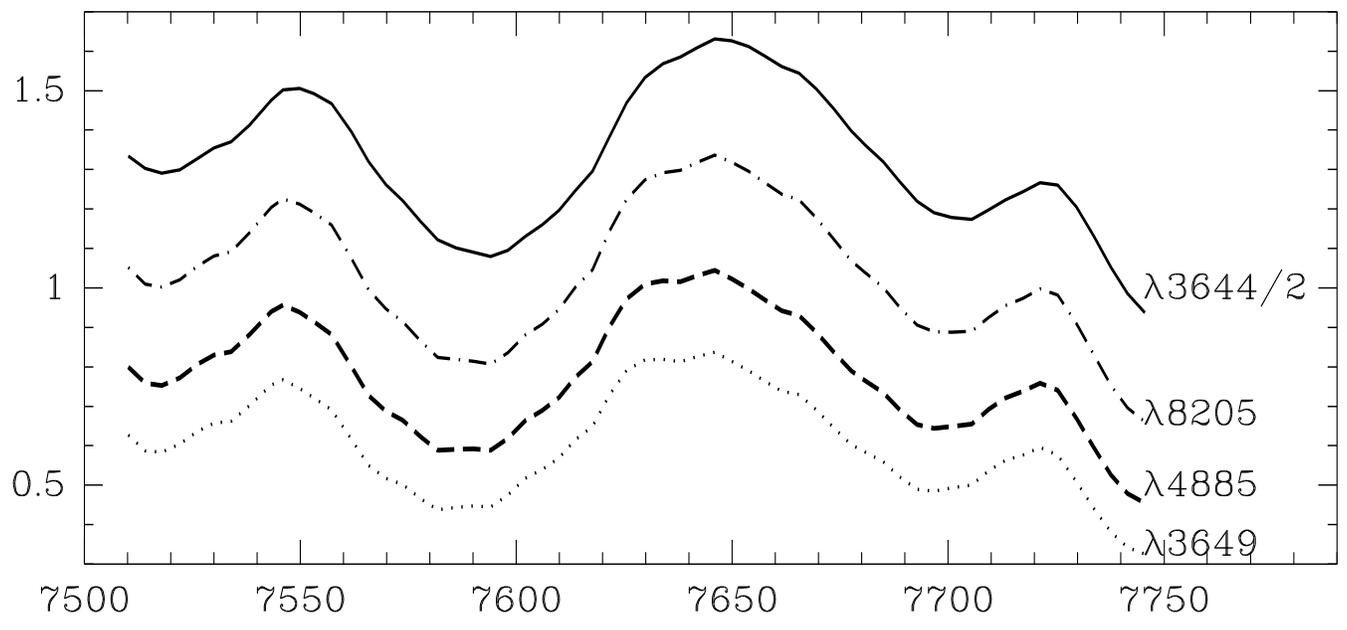
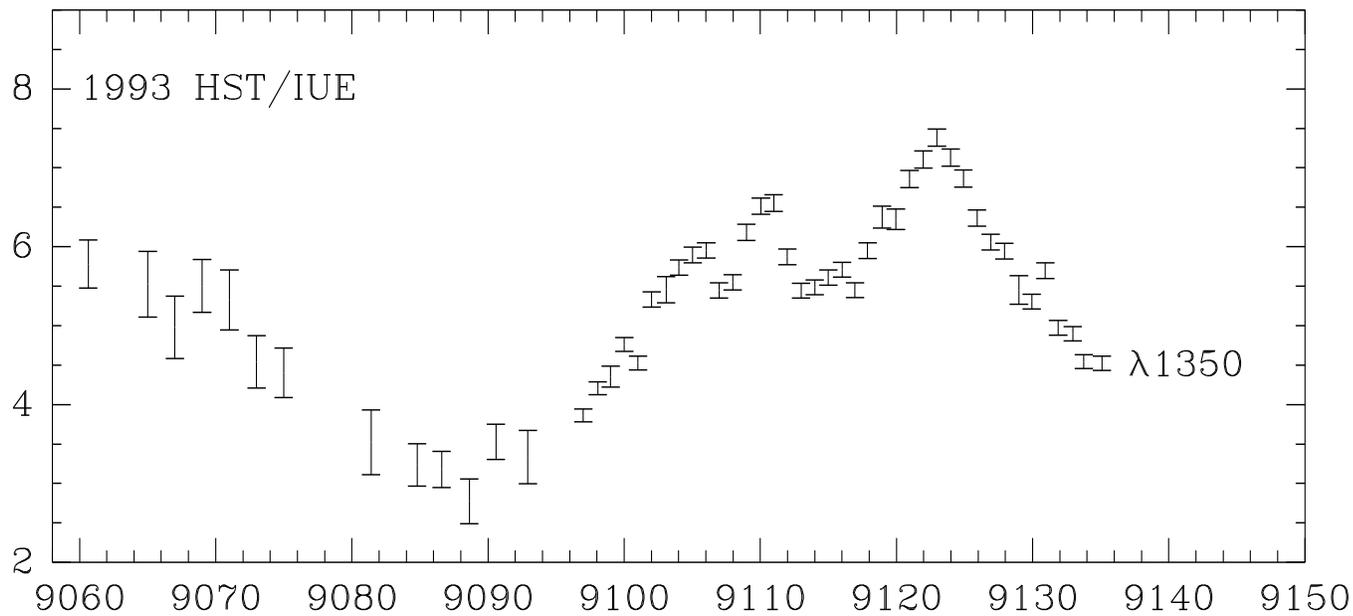
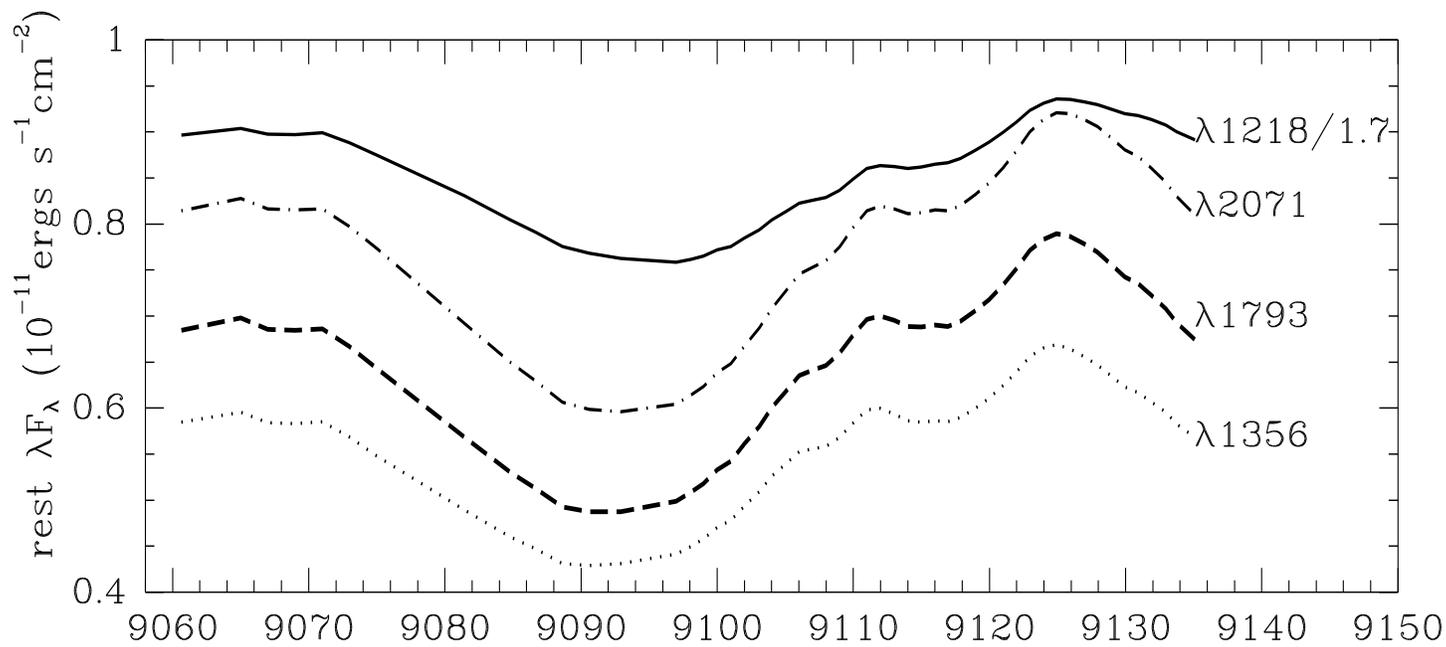
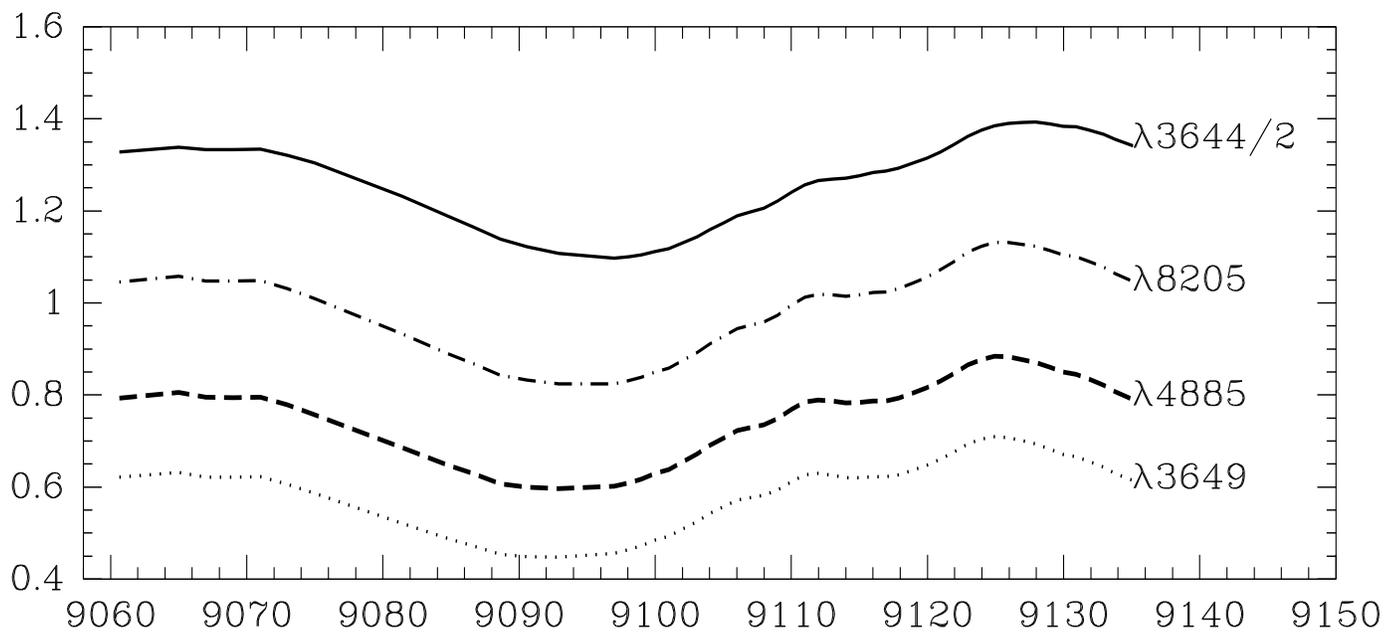
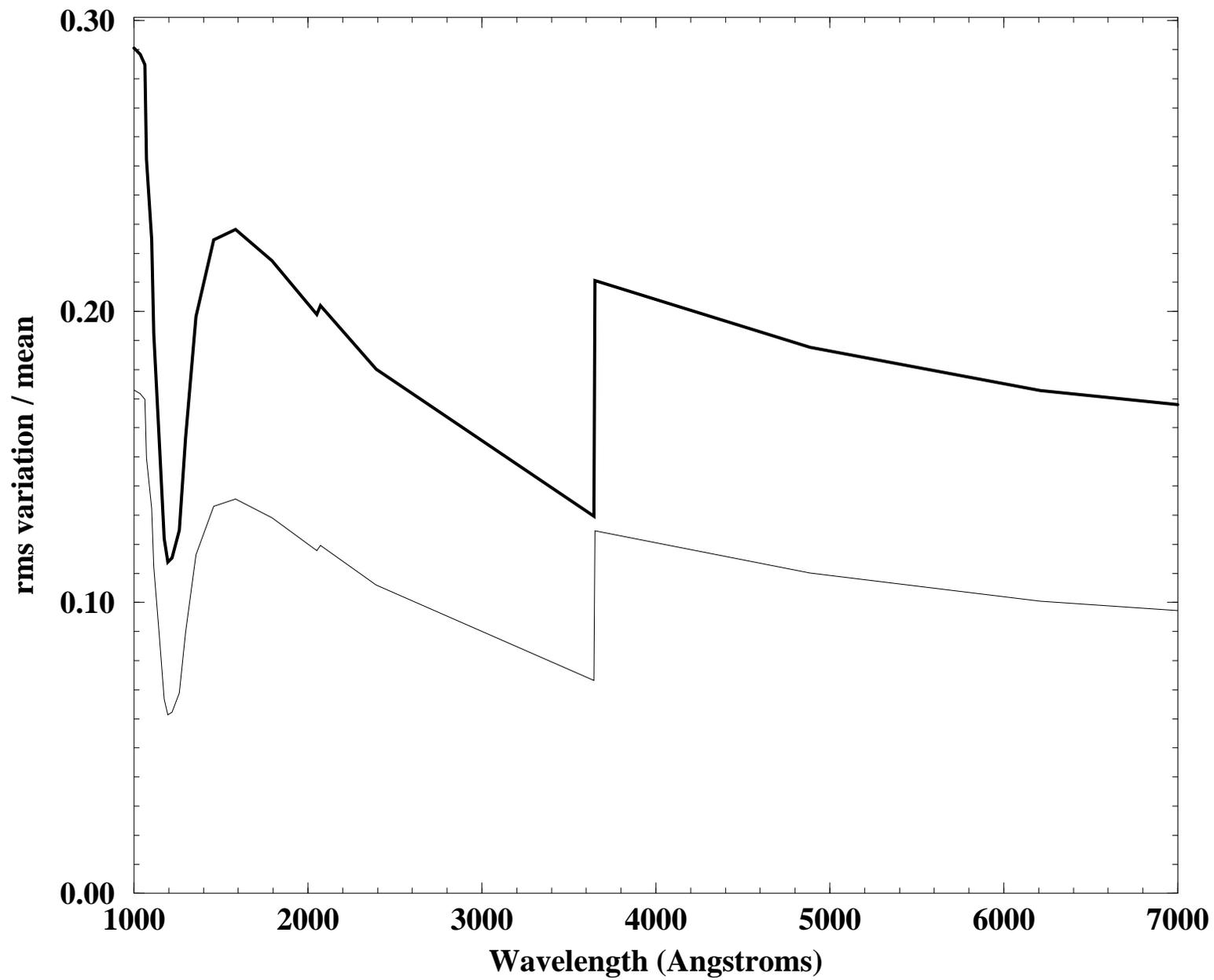
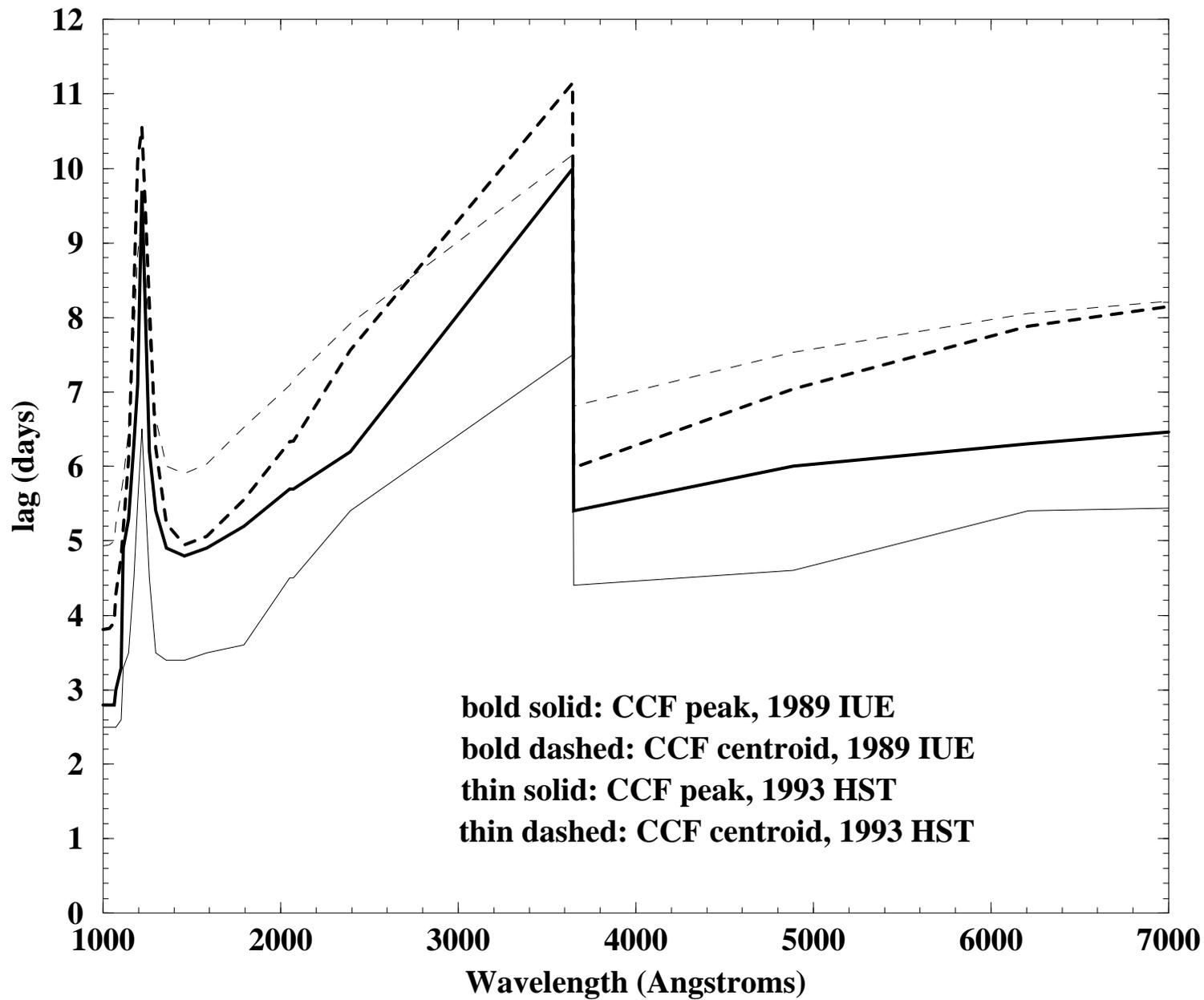
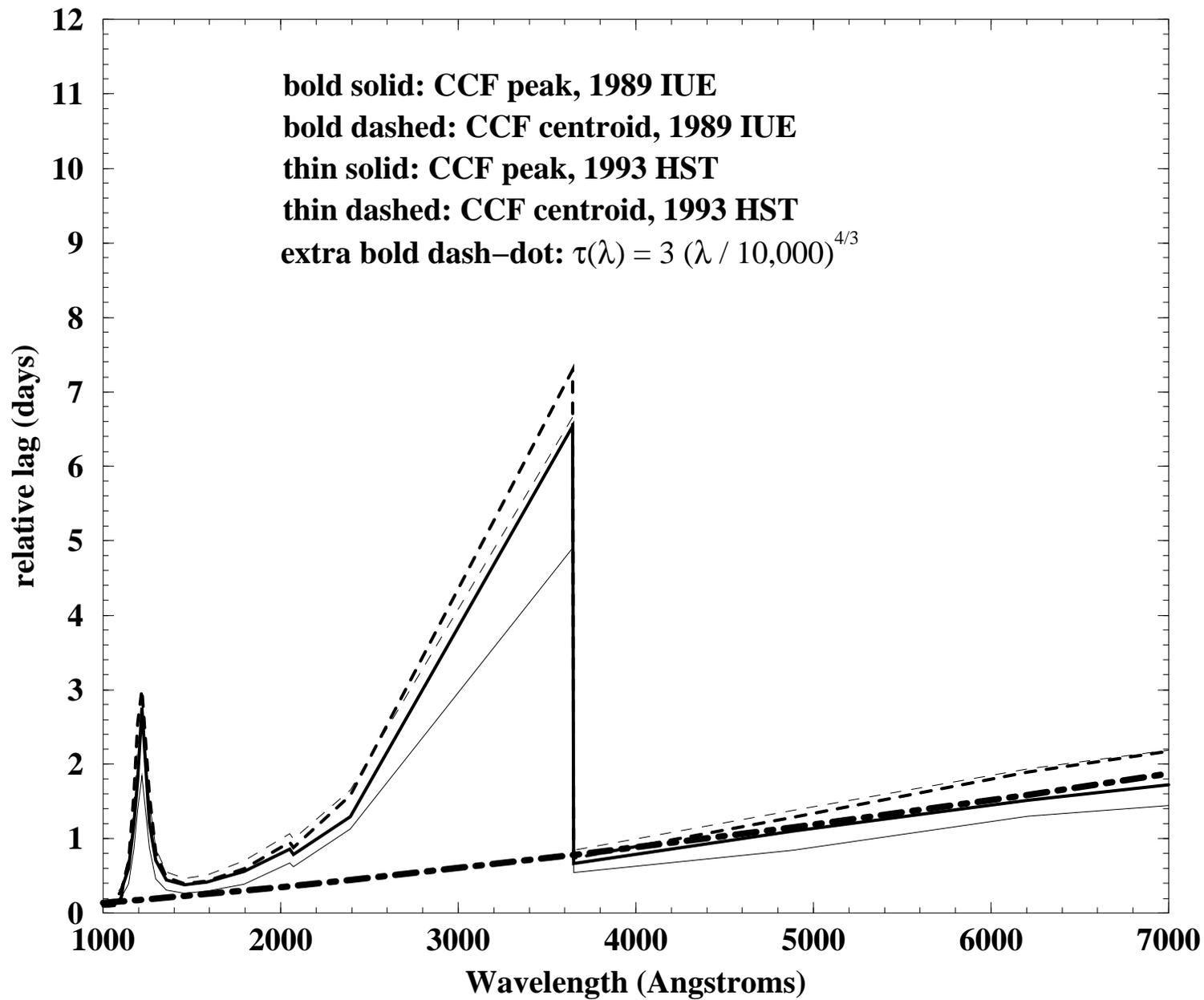
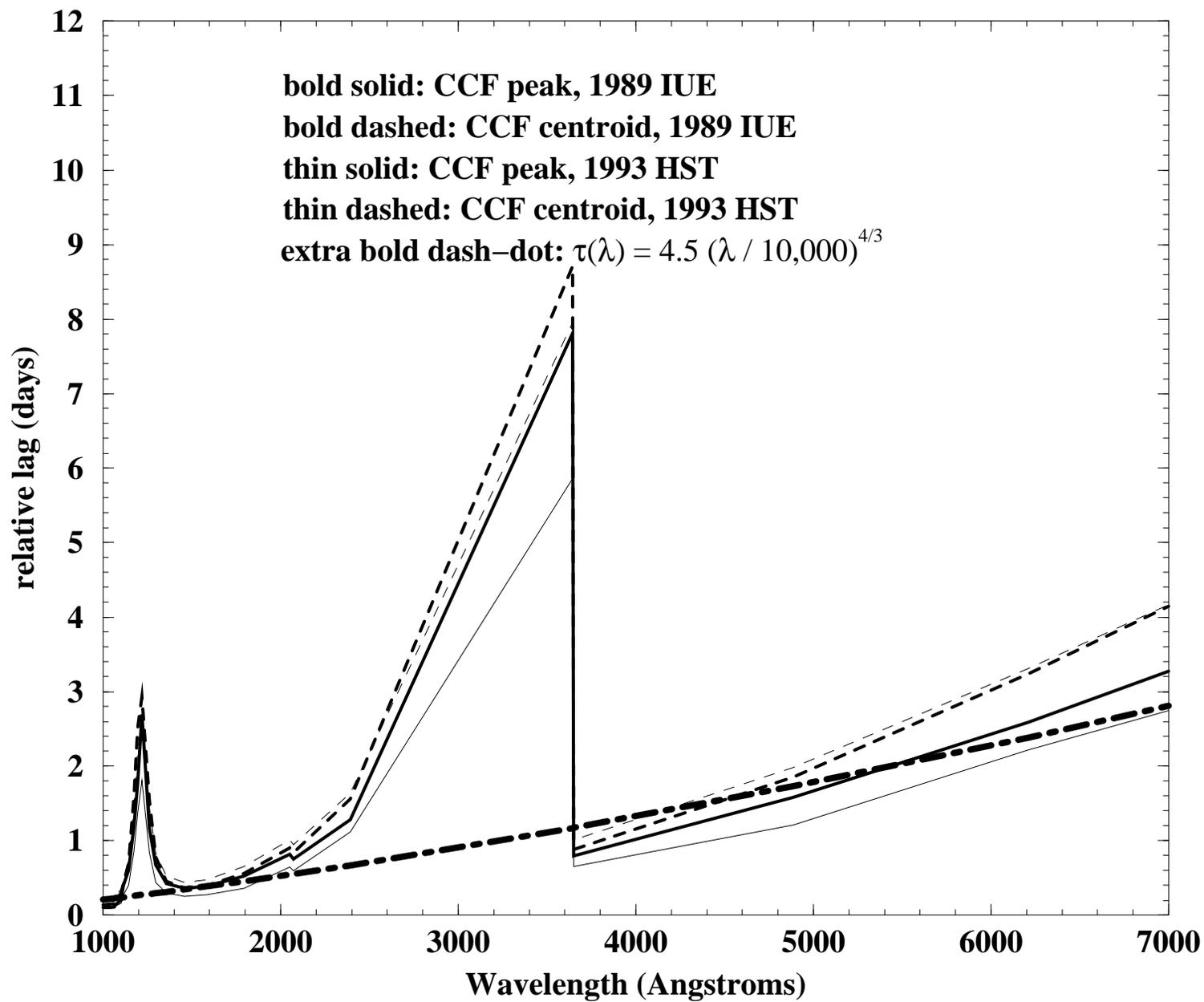
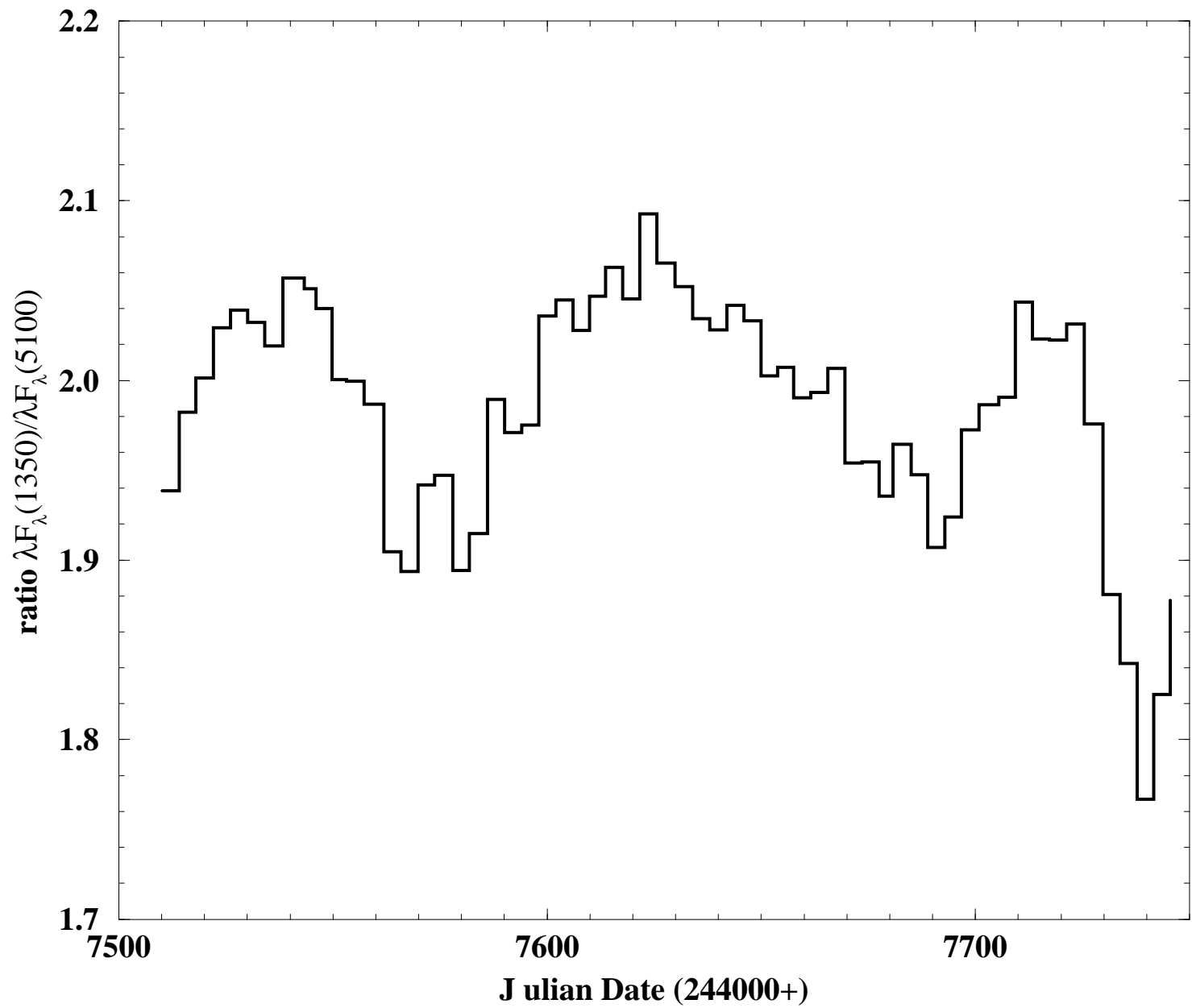
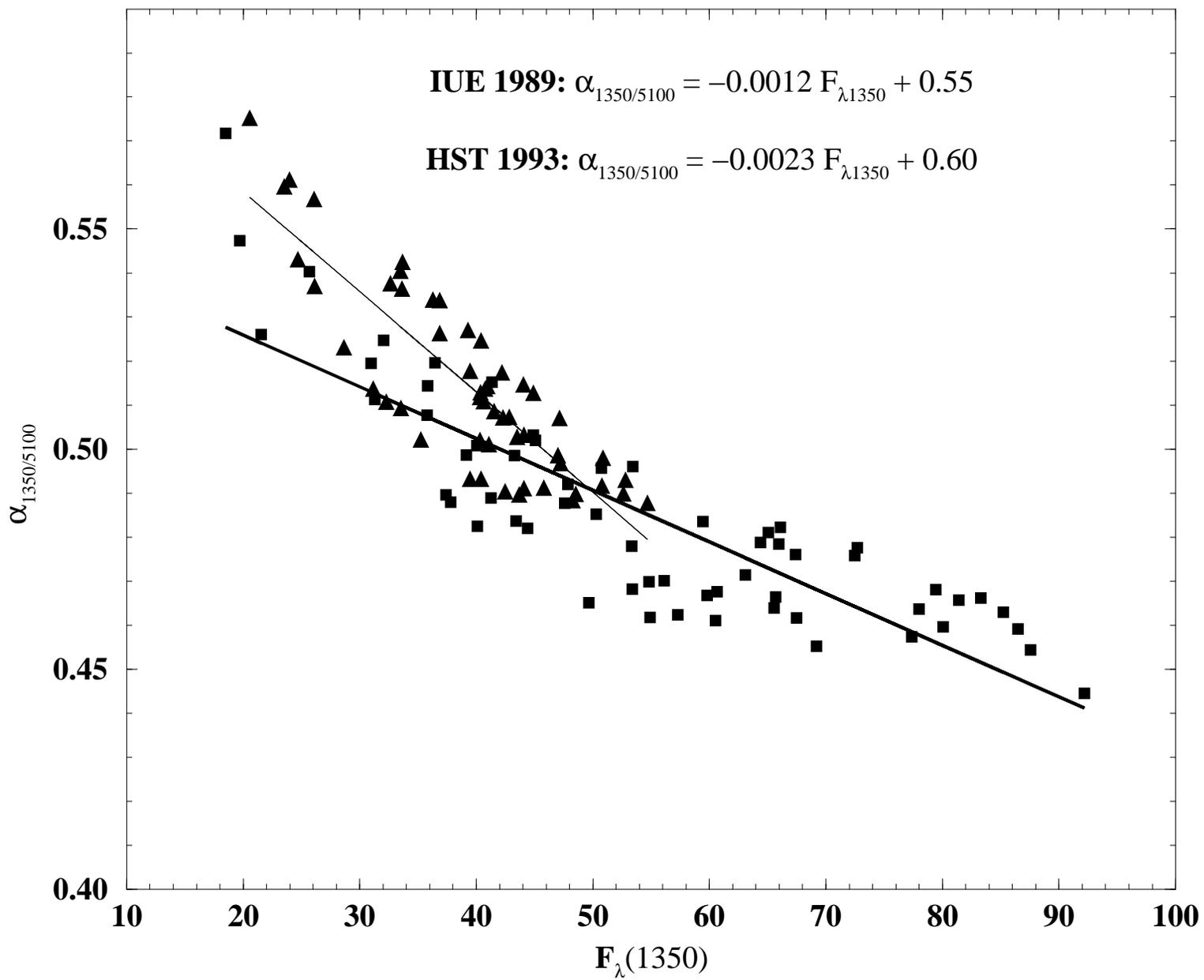
\end{document}